\documentclass[
  aps,
  prd,
  reprint,
  superscriptaddress,
  nofootinbib,
  amsmath,
  amssymb,
  floatfix
]{revtex4-2}

\usepackage{bm}
\usepackage{graphicx}
\usepackage{microtype}
\usepackage{xcolor}
\usepackage{hyperref}
\usepackage{orcidlink}

\graphicspath{{images/}}

\hypersetup{
  colorlinks=true,
  linkcolor=blue!55!black,
  citecolor=blue!55!black,
  urlcolor=blue!55!black,
  pdfauthor={Sidney Natzuka Junior, Carlos Augusto Domingues Zarro, Matheus dos Santos Soares},
  pdftitle={Field quantization in rotating frames: coordinate covariance and the circular-detector response},
  pdfsubject={Quantum field theory in rotating frames; Unruh-DeWitt detector response},
  pdfkeywords={quantum field theory, rotating frames, Unruh-DeWitt detector, Wightman function, Bogoliubov transformation, Trocheris-Takeno}
}

\newcommand{\dd}{\mathrm d}
\newcommand{\e}{\mathrm e}
\newcommand{\ii}{\mathrm i}
\newcommand{\M}{\mathrm M}
\newcommand{\RR}{\mathrm{rig}}
\newcommand{\TT}{\mathrm{TT}}
\newcommand{\order}[1]{\mathcal O\!\left(#1\right)}
\newcommand{\KG}[2]{\left(#1,#2\right)_{\mathrm{KG}}}
\newcommand{\Wight}{W^{+}}
\newcommand{\KTT}{\partial_t^{\TT}}
\begin{document}

\title{Field quantization in rotating frames: coordinate covariance and
the circular-detector response}

\author{Sidney Natzuka Junior\,\orcidlink{0009-0003-4756-6108}}
\email{sidney.natzuka@unesp.br}
\affiliation{Instituto de F\'isica Te\'orica, Universidade Estadual Paulista (UNESP), Rua Dr. Bento Teobaldo Ferraz 271, 01140-070 S\~ao Paulo, SP, Brazil}

\author{Carlos Augusto Domingues Zarro\,\orcidlink{0000-0002-4357-5168}}
\email{carlos.zarro@if.ufrj.br}
\affiliation{Instituto de F\'isica, Universidade Federal do Rio de Janeiro, Rua Milton Santos, 117, 21941-585, Cidade Universit\'{a}ria, Rio de Janeiro, RJ, Brazil}

\author{Matheus dos Santos Soares\,\orcidlink{0000-0001-5000-952X}}
\email{matheus.soares@cbpf.br}
\affiliation{Centro Brasileiro de Pesquisas F\'isicas, Rua Xavier Sigaud 150, 22290-180 Rio de Janeiro, RJ, Brazil}

\date{\today}

\begin{abstract}
We provide a unified analysis of scalar-field quantization and detector response in relativistic rotating frames, comparing rigid and Trocheris–Takeno (TT) descriptions in unbounded Minkowski spacetime. We distinguish three questions that are often conflated: \textbf{(i)} how a fixed quantum state is represented in rotating coordinates, \textbf{(ii)} whether rotating time evolution selects a global ground state, and \textbf{(iii)} how a rotating detector responds to the vacuum. Coordinate-transformed rigid and TT modes span the same positive-frequency subspace as inertial modes: their Bogoliubov beta coefficients vanish and their mode sums reproduce the Minkowski Wightman function. Neither rotating time flow selects a global scalar ground state, although for different reasons: the rigid generator is not globally timelike and is unbounded below, whereas TT time translation is not a Killing symmetry. Nevertheless, a circular Unruh–DeWitt detector in the Minkowski vacuum has a stationary, nonzero response. We relate this response to negative corotating frequencies, evaluate it after an explicit Hadamard subtraction, and show that it is nonthermal. Thus vanishing Bogoliubov mixing, the absence of a symmetry-selected rotating vacuum, and detector excitation are mutually consistent.
\end{abstract}

\maketitle

\section{Introduction}
\label{sec:introduction}

The particle content of a quantum field is tied to a choice of state and
a notion of time evolution, not to a coordinate label alone.  We
restrict attention to the vacuum case: every detector calculation below
uses the Minkowski vacuum, and we ask what a rotating description does,
and does not, change about that state.  The Unruh effect is the standard
illustration of the distinction between the field state, the observer's
trajectory, and the coordinates used to describe them
\cite{Fulling1973Nonuniqueness,Davies1975ScalarProduction,
Unruh1976BlackHole,CrispinoHiguchiMatsas2008Unruh}.  Consider a real
scalar field in the Minkowski vacuum and a
two-level Unruh--DeWitt detector coupled locally to that field.  If the
detector has constant proper acceleration \(a\), it excites even though
an inertial detector coupled to the same Minkowski-vacuum field does
not.  More precisely, the detector's excitation and decay rates obey
detailed balance with temperature
\[
 T_{\mathrm U}=\frac{a}{2\pi}
\]
in units \(c=\hbar=k_{\mathrm B}=1\).

There are two consistent descriptions.  In inertial coordinates one
evaluates the Minkowski (positive-frequency) Wightman function at two
points on the
accelerated worldline and Fourier transforms the result with respect
to detector proper time.  The resulting correlation function has the
imaginary-time periodicity required by the
Kubo--Martin--Schwinger (KMS) condition, the mathematical criterion for
thermal equilibrium with respect to a chosen time evolution; for a
stationary detector it implies thermal detailed balance
\cite{Sewell1982ThermalStates,FullingRuijsenaars1987}.
Alternatively, one may quantize in
the right Rindler wedge, the region \(X>|T|\) covered by uniformly
accelerated observers.  In that region the Lorentz-boost Killing
vector is timelike and generates translations of Rindler time, the
time coordinate followed by the uniformly accelerated observers.
Positive frequency with respect to this symmetry defines Rindler
modes, whose relation to Minkowski modes yields the same
Planck distribution.  Section~\ref{sec:unruh-control} states these
ingredients explicitly.

Uniform acceleration is a special case.  A timelike Killing field is not required for canonical quantization on a globally hyperbolic
spacetime, but a suitable stationary symmetry can select a preferred
positive-frequency splitting and a conserved Hamiltonian
\cite{Fulling1973Nonuniqueness,AshtekarMagnon1975,Kay1978,
Wald1994QFTCS,Sanders2013}.  Without
one, canonical quantization still proceeds from Cauchy data after a
compatible complex structure has been chosen.  This is an ordinary
Fock-space construction, but the geometry no longer selects a unique
choice, and different choices need not be unitarily equivalent
\cite{BirrellDavies1982QFCS,Wald1994QFTCS,
Radzikowski1996Microlocal,FewsterVerch2013Hadamard}.  This freedom to
choose a state must not be confused with coordinate freedom.  Once the
state, observable, detector trajectory, and switching are fixed,
coordinate covariance requires every chart and every complete mode
basis representing that same state to give the same detector response
\cite{BirrellDavies1982QFCS,Wald1994QFTCS,Takagi1986VacuumNoise,
CrispinoHiguchiMatsas2008Unruh,LoukoSatz2008Transition}.
Thus a calculation of a detector excitation (``clicking'') rate begins
by specifying the field state as well as the worldline and switching
function.

Rotation is a natural test of these distinctions.  Velocity and
acceleration are not parallel, the radius and angular velocity are
independent kinematic parameters, and sustained relativistic motion is
more plausible on a circular path than on an indefinitely long
linearly accelerated one.  The classical geometry of rotating disks is
directly relevant to field quantization: rigidity determines whether
the observer congruence remains timelike, while simultaneity and
foliation determine whether a proposed time coordinate supplies Cauchy
surfaces for initial data and the Klein--Gordon product.  These issues
were already central to relativistic discussions of rotating disks
\cite{Franklin1922Rotation,Gron1975RotatingDisk,
Herrera2000RotatingFrames,Post1967Sagnac}.  Rotation therefore
brings global complications: rigid motion reaches a speed-of-light
surface, whereas the Trocheris--Takeno construction avoids that surface
by assigning a radius-dependent angular velocity to the worldlines that
carry fixed spatial labels in its chart
\cite{Franklin1922Rotation,Trocheris1949Electrodynamics,
Takeno1952RotatingDisk}.

The preceding distinction between state, time evolution, and detector
trajectory becomes especially sharp for rotation.  The inertial and
Rindler descriptions illustrate how the same detector physics can be
organized either through a worldline correlation function or through a
symmetry-adapted quantization.  For rotating systems, however, the rigid
and Trocheris--Takeno (TT) time flows encounter different global
obstructions.  A rotating coordinate transformation must therefore be
distinguished both from vacuum selection and from restricting a chosen
state's two-point function to a detector worldline
\cite{BirrellDavies1982QFCS,Wald1994QFTCS,
LetawPfautsch1980RotatingCoordinates,
LetawPfautsch1981StationaryCoordinates}.

The question of vacuum selection is broader than any particular rotating
coordinate system.  For rigid rotation, coordinate time is generated by
a Killing field, but that field ceases to be timelike at the
speed-of-light cylinder and the scalar corotating Hamiltonian is
unbounded below on the full Minkowski space.  For TT rotation, the
worldlines at rest in the TT chart remain timelike at all radii, but the
associated time flow is not Killing and the constant-time surfaces do
not provide global Cauchy surfaces, as shown in
Sec.~\ref{sec:TT-obstruction}.  Consequently, neither construction
selects a distinct global scalar ground state throughout unbounded
Minkowski spacetime
\cite{Vilenkin1980RotatingSystem,DuffyOttewill2003RotatingThermal,
AmbrusWinstanley2021Rotation,LetawPfautsch1981StationaryCoordinates}.
This does not imply that the Minkowski vacuum cannot be represented in
rotating coordinates, that no other Hadamard states exist, or that
nonstationary canonical quantization is impossible; the precise scope of
the statement, and the two distinct mechanisms behind it, are set out at
the beginning of Sec.~\ref{sec:no-go} and established in
Secs.~\ref{sec:rigid-obstruction} and \ref{sec:TT-obstruction}.  

Two broader strands of the
rotating-field literature can appear to conflict.  Mode analyses may
find vanishing Bogoliubov beta coefficients between inertial and
rotating-coordinate bases, while detector analyses find that a
circularly moving detector becomes excited in the Minkowski vacuum
\cite{LetawPfautsch1980RotatingCoordinates,
DaviesDrayManogue1996RotatingVacuum,
DeLorenciEtAl2000RotatingDetector,BiermannEtAl2020CircularTemperature}.
Both statements can hold because they answer different questions.  A
beta coefficient compares the positive-frequency subspaces associated
with two mode decompositions, whereas a detector rate is the Fourier
transform of a two-point function restricted to the detector trajectory.
Along a circular orbit, modes of positive inertial frequency can have
negative corotating frequency \(\omega-m\Omega\), allowing detector
excitation without negative-frequency mixing between the complete mode
bases.  A detector click therefore does not diagnose the existence of a
distinct rotating vacuum.  

Furthermore, what a detector click diagnoses also depends on the
detector model.  The Unruh--DeWitt monopole used throughout this work
exchanges energy with the field and therefore responds to the Wightman
function restricted to its worldline, whereas a counting detector of
Glauber type is defined by normal ordering with respect to a chosen mode
decomposition, so that its response is tied more directly to the
positive-frequency splitting and hence to the Bogoliubov coefficients
\cite{Soares:2020qlw}.  The two statements are compatible, and which
question a given detector answers is fixed by its coupling.

The purpose of this work is to place these results in a single
consistent framework.  We analyze vacuum selection for rigid and TT
time evolution, construct the corresponding coordinate-transformed
modes, determine their normalization and Bogoliubov relations,
reconstruct the complete Wightman functions, and calculate the
circular-detector response in the Minkowski vacuum.  The resulting
picture sharpens distinctions used throughout quantum field theory in
noninertial settings
\cite{Wald1994QFTCS,LetawPfautsch1981StationaryCoordinates,
LoukoSatz2008Transition}: \textbf{(i)} a coordinate representation of a
state, \textbf{(ii)} a ground state selected by a time-evolution
symmetry, and \textbf{(iii)} the response of a localized noninertial
detector.

To implement this separation, we examine a free scalar field in
unbounded Minkowski spacetime.  Section~\ref{sec:stationarity} states the quantization
requirements and terminology used throughout, derives inertial
cylindrical modes, and uses Graf's addition theorem
\cite{Watson1944Bessel,DLMF} to recover the Cartesian Wightman function
in Sec.~\ref{sec:graf}; this is the template for the later rotating-mode
sums.  Section~\ref{sec:rotating-maps} introduces the rigid and
Trocheris--Takeno (TT) transformations as observer-adapted coordinates
in which the corresponding rotating observers have fixed spatial
 coordinates.  Section~\ref{sec:no-go} then shows that neither the rigid
 coordinate-time translation nor the TT coordinate-time translation
 selects a global scalar ground state through the corresponding rotating
 time-evolution symmetry in unbounded Minkowski space.  This is not the
claim that ``no vacuum state exists'' or that nonstationary canonical
quantization is impossible.  A cylindrical mirror is discussed there
because a boundary inside the speed-of-light cylinder defines a
 different, well-posed rotating boundary-value problem.  With those
 obstructions exposed, Sec.~\ref{sec:covariance} shows that the nullness of the Bogoliubov Beta-coefficients follows directly for modes defined as coordinate representatives
 of the inertial basis, and then verifies their admissibility,
 normalization, and reconstruction of the full Minkowski Wightman
 function.  A conclusion
convergent with ours, reached by a different, perturbative route, has
recently been obtained in Ref.~\cite{KuboniwaMameda2026Vacuum}.

Finally, we revisit a detector in circular motion.  Its excitation in the
Minkowski vacuum is nonzero but nonthermal
\cite{Letaw1981StationaryWorldLines,DaviesDrayManogue1996RotatingVacuum,
BellLeinaas1983Thermometers,Mueller1995RotatingSpectrum,
BiermannEtAl2020CircularTemperature,
GuttiKulkarniSriramkumar2011Rotating,ParryFewsterLouko2026}.
For the numerical plots we remove the universal coincident double pole
by inertial Hadamard subtraction and restore the known inertial rate.
This supplies a finite Fourier integral without changing
the state or hiding the singularity behind a finite regulator.  The
detector calculation, the subtraction, and the numerical evidence that
the resulting spectrum is not thermal are presented in
Sec.~\ref{sec:detector}. Finally, the conclusions are given in Sec.~\ref{sec:conclusion}.

\section{Stationarity and vacuum selection}
\label{sec:stationarity}

\subsection{Conventions and terminology}
\label{sec:conventions}

We work in \(3+1\) dimensions with metric signature \((+,-,-,-)\) and
units \(c=\hbar=k_{\mathrm B}=1\).  Partial derivatives are abbreviated
\(\partial_T\equiv\partial/\partial T\), and similarly for the other
coordinates.  We write \(\mathcal O\) for the spacetime region on which
the quantum field theory is formulated; a property is called
\emph{global on} \(\mathcal O\) when it holds at every point of
\(\mathcal O\).

Two-point functions appear in two roles, and we distinguish them
notationally throughout.  We write \(\Wight\) for the two-point function
of the field at two spacetime points, and \(\mathcal W^{+}\) for its
restriction to a single smooth timelike worldline, regarded as a
function of proper time along that worldline.  The restriction itself,
and the universal short-distance structure it inherits when the state is
Hadamard,\footnote{A state is \emph{Hadamard} when its
two-point function has the Hadamard short-distance form
\(\Wight\sim u/\sigma+v\ln\sigma+w\), with \(u,v,w\) smooth and
\(\sigma\) the squared geodesic separation, or equivalently when its
wavefront set obeys the microlocal spectrum condition
\cite{KayWald1991Bifurcate,Radzikowski1996Microlocal,
FewsterVerch2013Hadamard,BrunettiFredenhagenKohler1996}.} are defined where they are first needed, in
Sec.~\ref{sec:response-functional}.  This distinction between a
spacetime two-point function and its one-dimensional worldline
restriction is used explicitly in the detector response and pole
subtraction performed in Sec.~\ref{sec:detector}.

Coordinate maps, and the objects built from them, are named by a single
convention that is used from Sec.~\ref{sec:rotating-maps} onwards.  An
unsubscripted \(F\) always sends coordinate labels to the Minkowski
event that they label, \(X=F(x)\); the inverse map \(F^{-1}\) assigns
coordinate labels to a given event.  A subscript names the chart and
never a variable: \(F_{\RR}\) is the map from rigid rotating coordinates to inertial cylindrical coordinates, and \(F_{\TT}\) is the analogous map for the
Trocheris--Takeno coordinates. The map from
inertial to rigid rotating coordinates, is the inverse \(F^{-1}_{\RR}\), and the same follows for the TT map.
The same subscripts are carried by the metric functions, modes,
Wightman functions and Bogoliubov coefficients associated with each
chart.  In particular \(\RR\) abbreviates ``rigid'' and never denotes a
radius.

Three further conventions fix signs, and we collect them here rather
than at the point of first use.  The Klein--Gordon product
\(\KG{\cdot}{\cdot}\) of Eq.~\eqref{eq:KG-product} is antilinear in its
first argument and linear in its second, and the Bogoliubov
coefficients of Eq.~\eqref{eq:bogoliubov-overlaps} are defined with that
ordering.  Positive frequency is always fixed distributionally, through
the boundary value \(\Delta T\to\Delta T-\ii\epsilon\) with
\(\epsilon\to0^{+}\) taken after the relevant transform, as in
Eq.~\eqref{eq:cartesian-W-integral}.  Detector rates use the transform
convention of Eq.~\eqref{eq:stationary-rate}, with kernel
\(\e^{-\ii E\Delta s}\) and \(\Delta s=s-s'\), so that \(E>0\)
corresponds to excitation and \(E<0\) to de-excitation.

Finally, the combination \(\widetilde\omega=\omega-m\Omega\) that recurs
below, the Killing field \(\partial_T+\Omega\partial_\Phi\), and the
generator of rotating-frame time translations \(H-\Omega J_z\), are all
called \emph{corotating} (see Refs.
\cite{Vilenkin1980RotatingSystem,DuffyOttewill2003RotatingThermal,
AmbrusWinstanley2021Rotation}).

The remaining terminology concerns worldlines, and we keep two notions
strictly apart.  A worldline is called \emph{stationary} when it is an
orbit of a timelike Killing field, in the sense of the classification of
Refs.~\cite{Letaw1981StationaryWorldLines,Bunney2024}.  This is a
geometric property of the curve, and we never apply the word to a
coordinate system.  When the weaker, chart-dependent statement is
meant---that a worldline carries constant spatial coordinate
labels---we say that it is \emph{at rest in that chart}, and we speak of
the family of such curves as the worldlines at rest in the chart, rather
than as stationary coordinate lines.  Being at rest in a chart does not
imply stationarity, and in what follows the two notions come apart in
both directions: the rigid coordinate lines of Sec.~\ref{sec:rigid-map}
are at rest in the rigid chart at every radius but stationary only
inside the speed-of-light cylinder, whereas the Trocheris--Takeno
coordinate lines of Sec.~\ref{sec:TT-map} are at rest in their chart and
stationary at every radius, yet no single Killing field generates all of
them.  We reserve \emph{static} for the standard geometric condition of
a hypersurface-orthogonal timelike Killing field, again as a property of
a spacetime or of a Killing field and never of a chart.

\subsection{What a Killing field does---and does not---provide}
\label{sec:killing-role}

Whether a symmetry selects a global vacuum depends not only on a Killing vector
field but also on the spacetime region on which the field lives and if the
Cauchy problem on such region is posed
\cite{Kay1978,KayWald1991Bifurcate,Wald1994QFTCS,Kay1992Locality,
Sanders2013}.  This is why the comparison below distinguishes
unbounded Minkowski space, the Rindler wedge, and the interior of the
rigid speed-of-light cylinder.  The region \(\mathcal O\) may be a
proper---that is, strictly smaller---subregion of a larger spacetime.  For example, the right
Rindler wedge \(X>|T|\)
is a proper subset of Minkowski spacetime but is globally hyperbolic
when regarded as the spacetime on which Rindler quantization is
defined.  Specifying \(\mathcal O\) is therefore part of any claim about
a global vacuum.

For a real Klein--Gordon field, a quasifree\footnote{A state is
\emph{quasifree} when all of its \(n\)-point functions are determined by
its two-point function through the Wick (Gaussian) formula, so that
specifying \(\Wight\) specifies the state completely
\cite{Wald1994QFTCS,StreaterWightman2000PCT}.} Fock representation is
selected by a compatible complex structure on the classical solution
space, or equivalently by a positive-frequency subspace
\cite{AshtekarMagnon1975,Fulling1973Nonuniqueness}.  Here a \emph{complex structure} means a real-linear map \(J\) on the
space of real solutions with \(J^2=-1\), compatible with the symplectic
form; it encodes the splitting into positive- and negative-frequency
parts. A passive coordinate change merely relabels these solutions without
altering \(J\). It is therefore distinct from an active symplectic
transformation that changes the complex structure itself, whose unitary
implementability on Fock space is governed by Shale's theorem
\cite{Shale1962LinearSymmetries,BrunettiFredenhagenVerch2003}.

Each compatible complex structure determines a unique quasifree vacuum
state on the CCR algebra. To be physically acceptable, such a state must
be a normalized positive linear functional whose two-point distribution
satisfies the canonical commutation relations and obeys the Hadamard
condition. The Hadamard condition governs the short-distance singularity
structure: \(\Wight\sim u/\sigma+v\ln\sigma+w\), where \(u,v,w\) are
smooth functions and \(\sigma\) is the squared geodesic separation; the
equivalent microlocal spectrum condition is given in
Refs.~\cite{Radzikowski1996Microlocal,FewsterVerch2013Hadamard}.

Global hyperbolicity supplies a well-posed initial-value problem, and
canonical quantization then needs exactly two further choices: a Cauchy
surface \(\Sigma\subset\mathcal O\) on which the initial data are posed
and the symplectic form is evaluated, and a complex structure \(J\) on
the space of real solutions, compatible with that symplectic form, which
fixes the positive-frequency subspace and hence the Fock vacuum.  The
first choice is immaterial, because the symplectic form is independent
of the surface; the second is not, and it is the choice that the rest of
this section is about.

A timelike Killing field \(K\) adds something stronger, and it is worth
saying precisely what.  The field equation holds on all of
\(\mathcal O\) and is not itself modified by anything; what an isometry
does is act on its \emph{solutions}.  Because the flow
generated by \(K\) is an isometry, it commutes with \(\Box+\mu^2\), so
the coordinate change generated by the flow maps the space of
solutions onto itself and defines a one-parameter group of symplectic
transformations of that space.  If the state is invariant under the same
group, frequency with respect to the flow parameter is well defined and
conserved.
On the one-particle Hilbert space, this flow is represented by a unitary
group; its self-adjoint infinitesimal generator is the
\emph{one-particle generator}.  If that generator is bounded below,
positive \(K\)-frequency can select
a stationary ground state.  If \(K\) is absent, or if its generator is
unbounded below, this symmetry-based prescription fails. In this case, other
Hadamard states can still be chosen
\cite{Wald1994QFTCS,FewsterVerch2013Hadamard,KayWald1991Bifurcate}.

Time invariance alone is insufficient for thermodynamic equilibrium.  A
ground state has a nonnegative time-evolution generator in its
Gelfand--Naimark--Segal (GNS) representation, whereas a finite-temperature
KMS state satisfies imaginary-time strip analyticity and the KMS boundary
condition \cite{FullingRuijsenaars1987,Sewell1982ThermalStates}.
Thermodynamically, a state is passive if no net work can be extracted by
a cyclic perturbation, and complete passivity singles out ground and KMS
states \cite{PuszWoronowicz1978}.

For stationary linear fields on spacetimes satisfying the required
Cauchy-surface and operator hypotheses \cite{Kay1978,Sanders2013},
passive states obey the microlocal spectrum condition under the
assumptions of Ref.~\cite{SahlmannVerch2000}.  Thermodynamic stability
(ground/KMS) and ultraviolet regularity (Hadamard) are therefore related
but conceptually distinct.  This separation is needed below: the
rotating constructions fail to select a ground state for specific
symmetry or spectral reasons, while the coordinate-rewritten Minkowski
state remains Hadamard and the circular detector still has a stationary,
non-KMS response. This will be investigated throughout the text.

The distinctions needed below are therefore
\begin{enumerate}
 \item existence of a canonical quantization after a choice of complex
 structure;
 \item invariance of a specified state under a flow;
 \item existence of a ground state or a KMS state for that flow;
 \item the Hadamard property of the state; and
 \item stationarity, and possibly detailed
 balance,\footnote{\emph{Detailed balance} is the statement that the
 excitation and de-excitation rates of a two-level detector with gap
 \(E>0\) satisfy
 \(\dot{\mathcal F}(E)/\dot{\mathcal F}(-E)=\e^{-\beta E}\) with a
 single, gap-independent \(\beta\), which may then be read as an inverse
 temperature.  For a stationary correlator it is equivalent to the KMS
 condition along the worldline
 \cite{FullingRuijsenaars1987,Sewell1982ThermalStates,
 Takagi1986VacuumNoise}.  A gap-dependent ratio, such as the one
 obtained for circular motion in Sec.~\ref{sec:not-thermal}, is
 precisely the failure of detailed balance.} of a detector on
 one orbit.
\end{enumerate}
The ground-state assertion requires a generator with nonnegative
spectrum, and the KMS assertion requires its own analytic condition
\cite{FullingRuijsenaars1987,Sewell1982ThermalStates,Kay1978,
Sanders2013}. 
Constancy of a detector's spatial coordinate labels establishes neither:
it proves neither the existence of a ground or KMS state (item 3),
nor the detailed balance in item 5.

\subsection{Revisiting the inertial vacuum}
\label{sec:inertial-primer}

Let us consider Minkowski spacetime in Cartesian coordinates.  Using
inertial cylindrical coordinates \(X=(T,r,\Phi,z)\), the
Minkowski metric reads
\begin{equation}
 \dd s^2=\dd T^2-\dd r^2-r^2\dd\Phi^2-\dd z^2,
 \label{eq:inertial-metric}
\end{equation}
with \(\Phi\sim\Phi+2\pi\).  The free field of mass \(\mu\geq0\)
satisfies
\begin{equation}
 \left[
 \partial_T^2-\frac1r\partial_r(r\partial_r)
 -\frac1{r^2}\partial_\Phi^2-\partial_z^2+\mu^2
 \right]\phi=0.
 \label{eq:inertial-KG}
\end{equation}
Canonical quantization proceeds in the standard way
\cite{BirrellDavies1982QFCS,Wald1994QFTCS}, and it is convenient to
separate its two steps.  The first step is purely classical: one
equips the space of solutions of Eq.~\eqref{eq:inertial-KG} with the
Klein--Gordon product
\begin{equation}
 \KG{f}{g}
 =\ii\int_\Sigma\dd\Sigma^a
 \left(f^*\nabla_a g-g\nabla_a f^*\right).
 \label{eq:KG-product}
\end{equation}
Here \(\nabla_a\) is the covariant derivative with a Levi-Civita connection of the metric
\eqref{eq:inertial-metric}; acting on the scalars \(f\) and
\(g\) it reduces to the partial derivative, \(\nabla_af=\partial_af\),
so that all of the geometry in Eq.~\eqref{eq:KG-product} sits in the
surface element.  That surface element is \(\dd\Sigma^a=n^a\dd\Sigma\), with
\(\Sigma\) any Cauchy surface, \(n^a\) its future-directed unit normal
and \(\dd\Sigma\) the induced volume element.  For solutions of
Eq.~\eqref{eq:inertial-KG}, current conservation makes the value
independent of the chosen surface
\cite{BirrellDavies1982QFCS,Wald1994QFTCS}.  Because the integrand in
Eq.~\eqref{eq:KG-product} is built covariantly, the value of the product
is a number attached to a pair of solutions and not to the coordinates
used to write them; this elementary observation is paramount in
Sec.~\ref{sec:covariance}.

The second step introduces the quantum theory, and it is where a choice
must be made.  One selects a complete set \(\{v_\lambda\}\) of solutions
of Eq.~\eqref{eq:inertial-KG} obeying
\(\KG{v_\lambda}{v_{\lambda'}}=\delta_{\lambda\lambda'}\) and
\(\KG{v_\lambda}{v^{*}_{\lambda'}}=0\), and declares its span to be the
positive-frequency subspace; by the discussion of
Sec.~\ref{sec:killing-role} this is exactly a choice of compatible
complex structure.  Expanding the field operator in
\(\{v_\lambda,v^{*}_\lambda\}\), the expansion coefficients become
operators \(a_\lambda\) and \(a^{\dagger}_\lambda\) whose commutation
relations follow from the canonical ones.  The Hilbert space is the Fock
space built on the unique normalized vector \(|0\rangle\) annihilated by
every \(a_\lambda\), obtained by acting on it with the
\(a^{\dagger}_\lambda\); the vacuum is therefore defined by the chosen
set of solutions and not by the field equation alone.  A different admissible choice
of \(\{v_\lambda\}\), one not obtained from the first by a
transformation that mixes only the \(v_\lambda\) among themselves,
defines a different positive-frequency subspace and hence a different
vacuum.  Whether the two Fock representations are then unitarily
equivalent is the question measured by the Bogoliubov coefficients of
Sec.~\ref{sec:covariance}.

In Cartesian coordinates the normalized positive-frequency solutions of
Eq.~\eqref{eq:inertial-KG} are
\begin{equation}
 v_{\bm p}(X)=
 \frac{\exp(-\ii\omega_{\bm p}T+\ii\bm p\cdot\bm x)}
 {\sqrt{2\omega_{\bm p}(2\pi)^3}},
 \qquad
 \omega_{\bm p}=\sqrt{\bm p^2+\mu^2},
 \label{eq:cartesian-modes}
\end{equation}
where \(\bm p\) is the linear momentum three-vector.  The prefactor
\([2\omega_{\bm p}(2\pi)^3]^{-1/2}\) is fixed by demanding
orthonormality with respect to Eq.~\eqref{eq:KG-product}.  On a surface
of constant \(T\) the surface element is
\(\dd\Sigma^{a}=(\partial_T)^{a}\dd^{3}\bm x\) and
\(\partial_Tv_{\bm p}=-\ii\omega_{\bm p}v_{\bm p}\), so
Eq.~\eqref{eq:KG-product} collapses to
\((\omega_{\bm p}+\omega_{\bm p'})\int\dd^{3}\bm x\,
v^{*}_{\bm p}v_{\bm p'}\).  The remaining plane-wave integral gives
\((2\pi)^{3}\delta^{3}(\bm p-\bm p')\), the prefactors supply
\([2\omega_{\bm p}(2\pi)^{3}]^{-1}\) on the support of the delta, and
the two factors of \(2\omega_{\bm p}\) cancel.  For the mixed product
the same steps leave a coefficient proportional to
\(\omega_{\bm p'}-\omega_{\bm p}\) multiplying
\(\delta^{3}(\bm p+\bm p')\), and \(\omega_{\bm p}\) is even in
\(\bm p\), so that coefficient vanishes on the support of the delta.
Hence
\begin{align}
 \KG{v_{\bm p}}{v_{\bm p'}}&=\delta^3(\bm p-\bm p'),\nonumber\\
 \KG{v_{\bm p}}{v_{\bm p'}^*}&=0.
 \label{eq:cartesian-normalization}
\end{align}
The quantized real field is therefore
\begin{equation}
 \phi(X)=\int\dd^3\bm p\,
 \left[a_{\bm p}v_{\bm p}(X)
 +a_{\bm p}^{\dagger}v_{\bm p}^{*}(X)\right],
 \label{eq:cartesian-field-expansion}
\end{equation}
where \(a_{\bm p}\) and \(a_{\bm p}^{\dagger}\) are annihilation and
creation operators satisfying
\begin{equation}
 [a_{\bm p},a_{\bm p'}^{\dagger}]
 =\delta^3(\bm p-\bm p'),
 \qquad [a_{\bm p},a_{\bm p'}]=0.
 \label{eq:cartesian-commutators}
\end{equation}
These operators define the Minkowski vacuum by
\(a_{\bm p}|0_{\M}\rangle=0\) for every \(\bm p\).

For a state represented by a vector \(|\Psi\rangle\), the Wightman
function is the unordered and unsubtracted expectation value of a
product of two field operators,
\begin{equation}
 \Wight_\Psi(X,X')
 =\langle\Psi|\phi(X)\phi(X')|\Psi\rangle.
 \label{eq:W-definition}
\end{equation}
It measures the field correlation between two spacetime points.  The superscript \(+\)
records that in the Minkowski vacuum the Fourier transform of
Eq.~\eqref{eq:W-definition} with respect to \(X-X'\) is supported on the
positive mass shell \(p^{0}>0\), which is why it is called the
\emph{positive-frequency} Wightman function.  For a
quasifree state this two-point function determines all higher correlation functions, so that
it alone fixes the theory
\cite{StreaterWightman2000PCT,Wald1994QFTCS}; its
restrictions and derivatives provide the basic input for detector
responses and renormalized local observables.  In the Minkowski vacuum,
the Cartesian mode expansion in the basis
\eqref{eq:cartesian-modes} gives
\begin{equation}
 \Wight_{\M,\mu}(X,X')
 =\int\frac{\dd^3\bm p}{(2\pi)^3\,2\omega_{\bm p}}\,
 \e^{-\ii\omega_{\bm p}(\Delta T-\ii\epsilon)
 +\ii\bm p\cdot\Delta\bm x}.
 \label{eq:cartesian-W-integral}
\end{equation}
Here \(\epsilon>0\).  The prescription
\(\Delta T\to\Delta T-\ii\epsilon\), followed by
\(\epsilon\to0^+\), makes the momentum integral convergent and fixes
the positive-frequency boundary value
\cite{BirrellDavies1982QFCS,Wald1994QFTCS}.  In the detector rate
integrals of Sec.~\ref{sec:detector} it also fixes on which side of the
real axis the poles lie, and therefore which contour contributes to the
excitation or the de-excitation of an Unruh--DeWitt detector.

Define
\begin{equation}
 \rho_\epsilon
 =-(\Delta T-\ii\epsilon)^2+|\Delta\bm x|^2.
 \label{eq:rho-inertial}
\end{equation}
Direct evaluation of Eq.~\eqref{eq:cartesian-W-integral}, presented first
in Appendix~\ref{app:graf}, gives
\begin{align}
 \Wight_{\M,0}(X,X')&=\frac{1}{4\pi^2\rho_\epsilon},
 \label{eq:massless-W}\\
 \Wight_{\M,\mu}(X,X')&=
 \frac{\mu}{4\pi^2}
 \frac{K_1(\mu\sqrt{\rho_\epsilon})}{\sqrt{\rho_\epsilon}},
 \qquad\mu>0.
 \label{eq:massive-W}
\end{align}

Separating Eq.~\eqref{eq:inertial-KG} in the cylindrical coordinates
\((T,r,\Phi,z)\), rather than transforming
Eq.~\eqref{eq:cartesian-modes}, gives the regular cylindrical modes
\begin{align}
 v_{qmk}(X)
 &=\frac1{2\pi}\sqrt{\frac{q}{2\omega}}\,
 J_m(qr)\e^{\ii m\Phi+\ii kz-\ii\omega T},
 \nonumber\\
 \omega&=\sqrt{q^2+k^2+\mu^2},
 \label{eq:cylindrical-modes}
\end{align}
where \(J_{m}(z)\) is the Bessel function of the first kind of order
\(m\), and \(q>0\), \(k\in\mathbb R\), \(m\in\mathbb Z\).  As in
Eq.~\eqref{eq:cartesian-normalization}, the prefactor is fixed by
requiring orthonormality with respect to the Klein--Gordon product
\eqref{eq:KG-product}.  Evaluated on a surface of constant \(T\), that
product factorizes into three integrals.  The angular and axial ones are
elementary,
\begin{align}
 \frac1{2\pi}\int_0^{2\pi}\dd\Phi\,\e^{\ii(m-m')\Phi}
 &=\delta_{mm'},\nonumber\\
 \frac1{2\pi}\int_{-\infty}^{\infty}\dd z\,\e^{\ii(k-k')z}
 &=\delta(k-k'),
 \label{eq:angular-axial-orthogonality}
\end{align}
while the radial one is the closure relation for Bessel functions of the
first kind \cite{Watson1944Bessel,DLMF},
\begin{equation}
 \int_0^\infty\dd r\,r\,J_m(qr)J_m(q'r)
 =\frac{\delta(q-q')}{q}.
 \label{eq:bessel-closure}
\end{equation}
Together these give
\begin{align}
 \KG{v_{qmk}}{v_{q'm'k'}}
 &=\delta(q-q')\delta_{mm'}\delta(k-k'),\nonumber\\
 \KG{v_{qmk}}{v_{q'm'k'}^*}&=0.
 \label{eq:cylindrical-normalization}
\end{align}
Notice that no new state has been introduced:
Eqs.~\eqref{eq:cartesian-modes} and
\eqref{eq:cylindrical-modes} are two bases of the same
positive-inertial-frequency solution space, a point already emphasized
in the early analyses of rotating and accelerated observers
\cite{DenardoPercacci1978RotatingObserver,
LetawPfautsch1980RotatingCoordinates,Letaw1981StationaryWorldLines}.

\subsection{Cartesian and cylindrical mode equivalence}
\label{sec:graf}

We now calculate the Minkowski-vacuum Wightman function in the
cylindrical basis and verify that it reproduces the Cartesian result
\eqref{eq:cartesian-W-integral}.  Using the cylindrical modes
\eqref{eq:cylindrical-modes} in Eq.~\eqref{eq:W-definition}, in place of
the plane-wave basis \eqref{eq:cartesian-modes}, gives
\begin{align}
 \Wight_{\M,\mu}(X,X')
 &=
 \sum_{m=-\infty}^{\infty}
 \int_0^\infty\dd q
 \int_{-\infty}^{\infty}\dd k\,
 \frac{qJ_m(qr)J_m(qr')}{8\pi^2\omega}
 \nonumber\\
 &\quad\times
 \exp\!\left[
 \ii m\Delta\Phi+\ii k\Delta z
 -\ii\omega(\Delta T-\ii\epsilon)
 \right].
 \label{eq:inertial-mode-sum}
\end{align}
The azimuthal sum is performed with Graf's addition theorem
\cite{Watson1944Bessel,DLMF}, which yields
\begin{equation}
 \sum_{m=-\infty}^{\infty}J_m(qr)J_m(qr')
 \e^{\ii m\Delta\Phi}
 =J_0(qd_\perp),
 \label{eq:graf}
\end{equation}
where
\[
 d_\perp^2=r^2+r'^2-2rr'\cos\Delta\Phi.
\]
Using Eq.~\eqref{eq:graf} in Eq.~\eqref{eq:inertial-mode-sum}, the
remaining integrals coincide with the
Cartesian three-momentum representation after the transverse momentum
angle has been integrated, as shown explicitly in
Appendix~\ref{app:graf}.  The step that closes the comparison is the
elementary identity
\begin{equation}
 |\Delta\bm x|^2=d_\perp^2+(\Delta z)^2,
 \label{eq:transverse-identity}
\end{equation}
where \(d_\perp\), by its definition above, is the magnitude of the
separation projected onto the \(x\)-\(y\) plane.  With
Eq.~\eqref{eq:transverse-identity}, the argument \(\rho_\epsilon\) of
Eqs.~\eqref{eq:massless-W} and \eqref{eq:massive-W} is reproduced term
by term, so the cylindrical sum
\eqref{eq:inertial-mode-sum} reproduces exactly the Cartesian
Wightman functions \eqref{eq:massless-W} and
\eqref{eq:massive-W}.  This calculation shows why changing
from Cartesian plane waves to cylindrical Bessel modes does not change
the state; the same reasoning will be used for the rotating
coordinates in Sec.~\ref{sec:covariance}.

\subsection{Uniform acceleration as a control case}
\label{sec:unruh-control}

The results collected in this subsection are standard; we recall them
only to fix notation and to have a well-understood case against which
the rotating calculations of
Secs.~\ref{sec:no-go}--\ref{sec:detector} can be compared.  Detailed
treatments may be found in
Refs.~\cite{CrispinoHiguchiMatsas2008Unruh,Takagi1986VacuumNoise,
BirrellDavies1982QFCS}.

A uniformly accelerated worldline in the \(X\) direction can be
written
\begin{equation}
 T(s)=a^{-1}\sinh(as),\qquad
 X(s)=a^{-1}\cosh(as),
 \label{eq:accelerated-worldline}
\end{equation}
where \(s\) is proper time.  This trajectory is an orbit of the
Lorentz-boost Killing field
\begin{equation}
 K_{\mathrm B}=a\left(X\partial_T+T\partial_X\right).
 \label{eq:boost-Killing}
\end{equation}
Indeed, on Eq.~\eqref{eq:accelerated-worldline},
\begin{equation}
 \left.K_{\mathrm B}\right|_{X(s)}
 =\cosh(as)\partial_T+\sinh(as)\partial_X
 =\frac{\dd}{\dd s}.
 \label{eq:boost-tangent}
\end{equation}
Its norm \(K_{\mathrm B}^2=a^2(X^2-T^2)\) is positive in the right
Rindler wedge \(X>|T|\).  This wedge is relevant because it is the
globally hyperbolic domain in which \(K_{\mathrm B}\) is timelike and
can define Rindler positive frequency.

Evaluating Eq.~\eqref{eq:massless-W} at two points of the worldline gives
\begin{equation}
 \mathcal W_{\mathrm{lin}}^+(\Delta s)
 =-\frac{a^2}{16\pi^2}
 \frac{1}{
 \sinh^2\!\left[\frac a2(\Delta s-\ii\epsilon)\right]},
 \qquad \epsilon\to0^+.
 \label{eq:unruh-W}
\end{equation}
Here and below the subscript ``lin'' indicates linear, that is uniform,
acceleration, and the script \(\mathcal W^{+}\) indicates the
restriction of the spacetime two-point function to a worldline, as
fixed in Sec.~\ref{sec:conventions}.  For an interaction
maintained for a long proper time, the response probability per unit
time is the transition rate \(\dot{\mathcal F}_{\mathrm{lin}}(E)\),
\cite{BirrellDavies1982QFCS,LoukoSatz2008Transition}, whose general
definition is set up in Sec.~\ref{sec:response-functional}.  It is the
Fourier transform of the restriction of the Wightman function to the
detector worldline:
\begin{align}
 \dot{\mathcal F}_{\mathrm{lin}}(E)
 &=\int_{-\infty}^{\infty}\dd\Delta s\,
 \e^{-\ii E\Delta s}\mathcal W_{\mathrm{lin}}^+(\Delta s)
 \nonumber\\
 &=\frac{E}{2\pi}\frac{1}{\e^{2\pi E/a}-1},
 \label{eq:unruh-rate}
\end{align}
with \(E>0\) for excitation and \(E<0\) for de-excitation.  It obeys
\begin{equation}
 \frac{\dot{\mathcal F}_{\mathrm{lin}}(E)}
 {\dot{\mathcal F}_{\mathrm{lin}}(-E)}
 =\e^{-2\pi E/a}.
 \label{eq:unruh-detailed-balance}
\end{equation}
By the principle of detailed balance,
Eq.~\eqref{eq:unruh-detailed-balance} is the statement that the detector
equilibrates at the temperature \(T_{\mathrm U}=a/2\pi\)
\cite{FullingRuijsenaars1987,Sewell1982ThermalStates}.  More
fundamentally, the two-point function is analytic in the strip
\(-\beta<\operatorname{Im}z<0\), and its boundary values satisfy
\begin{equation}
 \mathcal W^+(z-\ii\beta)=\mathcal W^+(-z),
 \qquad \beta=\frac{2\pi}{a}.
 \label{eq:KMS-condition}
\end{equation}
Equation~\eqref{eq:KMS-condition} is the KMS condition for a bosonic
thermal state \cite{FullingRuijsenaars1987}, and it implies the
detailed-balance relation \eqref{eq:unruh-detailed-balance}.

Rindler quantization reaches the same conclusion from the accelerated
observer's time translation.  The boost Killing field is timelike in
the wedge $X>|T|$ and its orbits include Eq.~\eqref{eq:accelerated-worldline}.
The standard Bogoliubov calculation gives
\cite{Fulling1973Nonuniqueness,Unruh1976BlackHole,
CrispinoHiguchiMatsas2008Unruh}
\[
 \langle0_{\M}|N_{\omega}^{\mathrm{Rindler}}|0_{\M}\rangle
 =\frac1{\e^{2\pi\omega/a}-1}.
\]
The wedge restriction, boost symmetry, state, and KMS property are all
essential.  A coordinate singularity or detector excitation alone
would not establish thermality.  The question of how far the
thermal reading of such a response can be pushed, and in what sense the
state is or is not thermal, is discussed at length in
Ref.~\cite{SciamaCandelasDeutsch1981}.

\section{Rotating coordinate transformations}
\label{sec:rotating-maps}

This section defines the two observer-adapted coordinate systems used in
the rest of the paper and identifies the worldlines represented by fixed
spatial labels in each chart.  Their causal and global properties decide
whether coordinate-time translation can support the stationary-state
construction tested in Sec.~\ref{sec:no-go}; the same maps are then used
in Sec.~\ref{sec:covariance} to rewrite the inertial modes and Wightman
function without changing the field state.

\subsection{Rigid rotation and the speed-of-light cylinder}
\label{sec:rigid-map}

Let \(y=(\tau,r,\varphi,z)\) denote coordinates that rotate with
constant angular velocity \(\Omega_{\RR}\) relative to the inertial
coordinates.  The rigid rotation coordinate change is
\begin{equation}
 F_{\RR}:\qquad
 T=\tau,\qquad
 \Phi=\varphi+\Omega_{\RR}\tau,\qquad
 r=r,\qquad z=z.
 \label{eq:rigid-transformation}
\end{equation}
The worldline of a point of fixed \((r,\varphi,z)\) is
\begin{equation}
 T=\tau,\qquad \Phi(\tau)=\varphi+\Omega_{\RR}\tau.
 \label{eq:rigid-worldline}
\end{equation}
In the inertial chart it circles with angular velocity
\(\dd\Phi/\dd T=\Omega_{\RR}\).
The line element becomes
\begin{equation}
 \dd s^2=(1-\Omega_{\RR}^2r^2)\dd\tau^2
 -2\Omega_{\RR}r^2\dd\tau\,\dd\varphi
 -\dd r^2-r^2\dd\varphi^2-\dd z^2.
 \label{eq:rigid-metric}
\end{equation}
The rigid coordinate-time vector
\begin{equation}
 K_{\RR}=\partial_\tau
 =\partial_T+\Omega_{\RR}\partial_\Phi
 \label{eq:rigid-Killing}
\end{equation}
is Killing, with
\[
 K_{\RR}^2=1-\Omega_{\RR}^2r^2.
\]
Its constant-spatial-coordinate orbits are timelike only for
\(r<|\Omega_{\RR}|^{-1}\).  As fixed in Sec.~\ref{sec:conventions}, the
subscript abbreviates ``rigid'' and never denotes a radius.

Two independent statements about Eq.~\eqref{eq:rigid-worldline} can now
be made, and the terminology of Sec.~\ref{sec:conventions} keeps them
apart.  First, the worldline is \emph{at rest in the rigid chart}: its
spatial labels \((r,\varphi,z)\) are constant.  This is a coordinate
statement and it holds at every radius.  Second, for
\(r<|\Omega_{\RR}|^{-1}\) the worldline is also \emph{stationary}: it is
an orbit of the Killing field \(K_{\RR}\) of
Eq.~\eqref{eq:rigid-Killing}, which is timelike there.  This is a
geometric statement, and it is the one that fails once
\(K_{\RR}\) ceases to be timelike.

We call the surface
\begin{equation}
 r=|\Omega_{\RR}|^{-1}
 \label{eq:light-cylinder}
\end{equation}
the \emph{speed-of-light cylinder}, and we use that name throughout.  By
definition it is the locus at which the corotating Killing field
\(K_{\RR}\) ceases to be timelike, equivalently the radius at which a
rigidly corotating observer would have to move at the speed of light.
It is \emph{not} a null hypersurface and it is not a horizon.  The name
refers to the causal character of \(K_{\RR}\) on the surface, not to the
causal character of the surface itself; the two are independent, and
conflating them is a source of confusion surrounding rotating charts.
We therefore separate them explicitly.

The first is the causal character of the Killing vector.  From
the norm above, \(K_{\RR}\) is timelike for \(r<|\Omega_{\RR}|^{-1}\),
null on Eq.~\eqref{eq:light-cylinder}, and spacelike for
\(r>|\Omega_{\RR}|^{-1}\).  This is what the name records.

The second is the causal character of the surface, which is governed by
its normal and not by \(K_{\RR}\).  Taking
\(f=r-|\Omega_{\RR}|^{-1}\), a normal covector is
\[
 n_a=\nabla_a f=(0,1,0,0),\qquad
 g^{ab}n_an_b=g^{rr}=-1.
\]
The vector \(n^a\) is not proportional to \(K_{\RR}^a\), and the value
\(g^{rr}=-1\) is independent of \(r\): in the metric
\eqref{eq:rigid-metric} the \(rr\) component decouples from the
\((\tau,\varphi)\) block.  The normal is therefore spacelike at every
radius, and every surface of fixed \(r\)---including
\(r=|\Omega_{\RR}|^{-1}\) itself, and every surface outside it---is a
timelike hypersurface.  The determinant
\(\det g_{\RR}=-r^2\) does not vanish there, and timelike curves can
cross it in either direction.

There is thus no tension between \(K_{\RR}^2<0\) beyond the cylinder and
the cylinder being timelike, but the point deserves an explicit
computation, because in Eq.~\eqref{eq:rigid-metric} every diagonal
component is negative once \(r>|\Omega_{\RR}|^{-1}\), which might be
read as a Euclidean signature.  It is not: signature cannot be read off
the diagonal when \(g_{\tau\varphi}\neq0\).  The \((\tau,\varphi)\)
block of Eq.~\eqref{eq:rigid-metric} has determinant
\begin{equation}
 g_{\tau\tau}g_{\varphi\varphi}-g_{\tau\varphi}^{2}
 =-r^{2}\left(1-\Omega_{\RR}^{2}r^{2}\right)-\Omega_{\RR}^{2}r^{4}
 =-r^{2},
 \label{eq:rigid-block-determinant}
\end{equation}
which is negative for every \(r>0\).  That block therefore always has
one positive and one negative eigenvalue, and together with
\(g_{rr}=g_{zz}=-1\) the signature is \((+,-,-,-)\) at every radius.
The chart is Lorentzian on both sides of the cylinder, and what
\(g_{\tau\tau}<0\) records is only that the particular direction
\(\partial_\tau\) has stopped being timelike. Note that even at the speed-of-light cylinder, where $g_{\tau \tau}=0$, the metric determinant is negative and therefore the coordinate transformation is well defined there, and for the whole spacetime except for the line $r=0$.

What fails outside the cylinder is accordingly neither the metric nor
the chart, but the interpretation of the coordinate lines as material,
rigidly corotating observers.  The coordinate transformation
\eqref{eq:rigid-transformation} itself remains nonsingular at
\(r=|\Omega_{\RR}|^{-1}\); it is the rigid observer congruence that
ceases to be timelike, while the spacetime remains flat and geodesically
complete throughout.  Physical and spectral conclusions must therefore
be based on the congruence and its generator, not on a coordinate singularity.

\subsection{The Trocheris--Takeno transformation}
\label{sec:TT-map}

Rigid rotation assigns the tangential speed
\(v=\Omega_{\RR}r\), so the worldlines at rest in its chart become null
at a finite radius; the circular orbits that remain timelike belong to the
classification of stationary worldlines in Minkowski spacetime given in
Refs.~\cite{Letaw1981StationaryWorldLines,Bunney2024}.  Relativistic
rotational transformations proposed by
Franklin, Trocheris, and Takeno instead accumulate rapidity with radius,
producing a subluminal but differentially rotating congruence
\cite{Franklin1922Rotation,Trocheris1949Electrodynamics,
Takeno1952RotatingDisk}.  Such transformations have been motivated by
different combinations of local Lorentz behavior, velocity-composition
properties, and rotating-disk geometry; they are not unique, and
modified Franklin/TT prescriptions have also been proposed
\cite{Herrera2000RotatingFrames,NouriZonozEtAl2014}.  We therefore fix
one precise TT \emph{orientation}---by which we mean a choice of sign
for the radial rapidity, that is, which of \(\pm\Omega_{\TT}\) is used
to build the transformation.  This choice says whether the congruence turns one way around the axis
or the other. 

The choice nevertheless carries a consequence that is easy to overlook,
and we return to it once the angular identification has been written
down.  The TT time coordinate is not built from the inertial time alone
but from the angle as well, so reversing the sense of rotation reverses
the sense in which TT time is displaced on going once around the circle. Thus, a consistent orientation is necessary for a correct calculation of the Bogoliubov coefficients related to this transformation, as discussed in Sec.~\ref{sec:TT-field}. Define
\begin{align}
 \eta(r)&=\Omega_{\TT}r,\nonumber\\
 C(r)&=\cosh\eta(r),\qquad S(r)=\sinh\eta(r).
 \label{eq:TT-CS}
\end{align}
Two maps are needed, and they are named by the convention fixed in
Sec.~\ref{sec:conventions}: an unsubscripted \(F\) sends coordinate
labels to the Minkowski event that they label.  The rigid map
\(F_{\RR}\) of Eq.~\eqref{eq:rigid-transformation} already follows it,
since it sends rigid labels to inertial ones.  Accordingly, the map from
inertial to TT labels is written \(F^{-1}_{\TT}\):
\begin{equation}
 F^{-1}_{\TT}:\qquad t=CT-rS\Phi,\qquad
 \theta=C\Phi-\frac{S}{r}T,
 \label{eq:TT-forward}
\end{equation}
and the map from TT labels back to inertial ones---the one actually
needed to express an inertial scalar in TT covering coordinates---is
\(F_{\TT}\):
\begin{equation}
 F_{\TT}:\qquad
 T=Ct+rS\theta,\qquad
 \Phi=C\theta+\frac{S}{r}t.
 \label{eq:TT-inverse}
\end{equation}
Here \(\eta\) is the rapidity of the local boost, so
\(\Omega_{\TT}=\dd\eta/\dd r\) is a constant radial rapidity gradient.
At fixed \(r\), Eq.~\eqref{eq:TT-forward} is a Lorentz boost of inertial
time \(T\) and tangential arclength \(r\Phi\).  Consequently, a curve
with fixed TT spatial labels maps to a circular inertial trajectory
whose tangential speed is the boost velocity \(\tanh\eta\), as is
verified explicitly below.

To define the functions in Eq.~\eqref{eq:TT-forward}, we work
first on the angular covering space, where
\(-\infty<\Phi<\infty\), just as polar coordinates may be unwrapped
before imposing periodicity.  For \(r>0\),
Eqs.~\eqref{eq:TT-forward} and \eqref{eq:TT-inverse} define the TT
covering coordinates \((t,r,\theta,z)\).  The axis \(r=0\),
where Eq.~\eqref{eq:TT-forward} degenerates, is excluded from this
coordinate description.  The passage back to the physical, punctured
Minkowski spacetime is carried out below, once the periodicity
\(\Phi\sim\Phi+2\pi\) has been imposed.

Differentiating Eq.~\eqref{eq:TT-inverse} and substituting into the
Minkowski line element gives the TT metric
\begin{align}
 \dd s^2={}&\dd t^2-(1+\mathfrak Q)\dd r^2-r^2\dd\theta^2-\dd z^2
 \nonumber\\
 &+2\mathfrak A\,\dd t\,\dd r
 +2\mathfrak B\,\dd r\,\dd\theta,
 \label{eq:TT-metric}
\end{align}
where
\begin{align}
 \mathfrak A&=\frac{S^2}{r}t+(CS+\eta)\theta,\nonumber\\
 \mathfrak B&=(CS-\eta)t+rS^2\theta,\nonumber\\
 \mathfrak Q&=
 \left(\eta^2-2\eta CS+S^2\right)\frac{t^2}{r^2}
 -4\eta S^2\frac{t\theta}{r}\nonumber\\
 &\quad-\left(\eta^2+2\eta CS+S^2\right)\theta^2.
 \label{eq:TT-metric-functions}
\end{align}
The explicit \(t\) and \(\theta\) dependence of
\(\mathfrak A\), \(\mathfrak B\), and \(\mathfrak Q\) already
anticipates that TT time translation is not a spacetime symmetry.

At fixed \((r,\theta,z)\),
\begin{equation}
 \frac{\dd T}{\dd t}=C,\qquad
 \frac{\dd\Phi}{\dd t}=\frac Sr,\qquad
 \dd s^2=\dd t^2.
 \label{eq:TT-orbit}
\end{equation}
Thus \(t\) is the proper time along a curve of constant TT spatial
coordinates, and
\begin{equation}
 v_{\TT}(r)=\tanh(\Omega_{\TT}r)<1
 \label{eq:TT-speed}
\end{equation}
at every finite radius.  The parameter \(\Omega_{\TT}\) is not the
angular velocity of every TT observer.  Their \emph{inertial angular
velocity}, by which we mean the kinematic
quantity \(\dd\Phi/\dd T\) measured in the inertial chart, is radius
dependent:
\begin{equation}
 \Omega_{\mathrm{phys}}^{\TT}(r)
 =\frac{\dd\Phi}{\dd T}
 =\frac{\dd\Phi/\dd t}{\dd T/\dd t}
 =\frac{\tanh(\Omega_{\TT}r)}r,
 \label{eq:TT-physical-Omega}
\end{equation}
where the last equality uses Eq.~\eqref{eq:TT-orbit}.  Each such curve
is therefore a circular orbit of the kind classified in
Refs.~\cite{Letaw1981StationaryWorldLines,Bunney2024}.  In the
terminology of Sec.~\ref{sec:conventions} such a curve is both at rest
in the TT covering chart, since its spatial labels are constant, and
stationary, since it is a circular Killing orbit.  Here the two notions
happen to agree, but only orbit by orbit: the Killing field involved is
a different one at each radius, which is the content of
Sec.~\ref{sec:TT-obstruction}.

The physical identification \(\Phi\sim\Phi+2\pi\), inherited from the
underlying Minkowski cylinder, induces
\begin{equation}
 (t,r,\theta,z)\sim
 (t-2\pi rS,r,\theta+2\pi C,z).
 \label{eq:TT-identification}
\end{equation}
Equation~\eqref{eq:TT-identification} identifies points along a helix in
the \((t,\theta)\) plane: one circuit in \(\Phi\) shifts \(t\) and
\(\theta\) together, rather than returning \(\theta\) to its initial
value at fixed \(t\).  Consequently, neither \(t\) nor \(\theta\)
descends to a single-valued global function on the physical spacetime.
The ranges of the covering coordinates are accordingly
\(t,\theta,z\in\mathbb R\) at each \(r>0\), and for brevity ``TT
coordinates'' below means these covering coordinates, with physical
scalar quantities required to be invariant under
Eq.~\eqref{eq:TT-identification}.

Note that no conical defect is
involved: the deficit angle vanishes, and the spacetime is the ordinary
Minkowski spacetime. This is the TT form of the familiar global synchronization obstruction
on a rotating disk
\cite{Gron1975RotatingDisk,Herrera2000RotatingFrames}.  The Sagnac time
gap is a holonomy around the
angular direction \cite{Post1967Sagnac}; geometrically, vorticity is the
curvature of the simultaneity connection and prevents global Einstein
synchronization \cite{Minguzzi2003}.  Alternative disk conventions can
move the discontinuity into a direction-dependent one-way light speed,
but cannot remove the global holonomy
\cite{Kassner2012,Koks2017}.  Equation~\eqref{eq:TT-identification}
is therefore part of the passage from the angular cover to the physical
spacetime, not an optional local coordinate convention.

The orientation chosen above is visible in
Eq.~\eqref{eq:TT-identification}.  Reversing it,
\(\Omega_{\TT}\to-\Omega_{\TT}\), leaves \(C\) unchanged and reverses
the sign of \(S\), so the identification becomes
\[
 (t,r,\theta,z)\sim(t+2\pi rS,r,\theta+2\pi C,z);
\]
one circuit in \(\Phi\) displaces the TT time coordinate in the opposite
sense, while the angular displacement, which involves \(C\) alone, is
unchanged.  This is the substantive content of the orientation.  It is not
only that the congruence turns the other way, which is the elementary
statement made above, but that \(t\) depends on the angle as well as on
the inertial time, through the term \(-rS\Phi\) of
Eq.~\eqref{eq:TT-forward}.

The magnitude of the displacement is the same for both orientations.
Since \(C\) is the Lorentz factor of the local boost and
\(S=C\,v_{\TT}(r)\), by Eqs.~\eqref{eq:TT-CS} and \eqref{eq:TT-speed},
\[
 |\Delta t|=2\pi rS=2\pi r\,C\,v_{\TT}(r),
\]
which is the Sagnac holonomy of the preceding paragraph written in TT
variables.  Its magnitude is physical and its sign is the orientation,
in the same way that the Sagnac time difference is itself odd under
reversal of the rotation orientation.

It must not, however, be read as a lapse of time experienced by any observer.
Equation~\eqref{eq:TT-identification} states that \((t,r,\theta,z)\) and
\((t-2\pi rS,r,\theta+2\pi C,z)\) are two labels for the \emph{same}
event: a single event carries infinitely many TT times, and no worldline
has been traversed between them.  If \(t\) is nevertheless treated as a
global time function and followed once around the circle, one appears to
return to an earlier moment, and the construction appears to admit
closed causal curves;  there are no such curves.  The appearance is an artifact of a
multivalued coordinate; the covering coordinates are single valued and
regular, and it is only their descent to the quotient that fails, while
the underlying spacetime is ordinary Minkowski spacetime, flat,
geodesically complete and free of closed causal curves.  Nor is the
failure repaired by rescaling the angle, since the same circuit
displaces \(\theta\) by \(2\pi C>2\pi\).  Neither covering coordinate is
separately periodic, and only the pair, related together as in
Eq.~\eqref{eq:TT-identification}, describes the physical spacetime.
This is why every physical quantity below is required to be invariant
under Eq.~\eqref{eq:TT-identification}.

\section{Obstructions to symmetry-selected rotating ground states}
\label{sec:no-go}

\noindent\textit{Scope of the statement.---}
Throughout this section we consider a free real bosonic Klein--Gordon
field on unbounded $3+1$-dimensional Minkowski spacetime, with
nonzero rotation and no material boundary.  By a state
\emph{selected by rotating time evolution} we mean a quasifree ground
state whose positive-frequency splitting is fixed by a single
time-translation flow tangent to the congruence of rotating worldlines, with an
induced conserved one-particle generator that is self-adjoint and
bounded below.  Under these assumptions, neither the rigid nor the TT
coordinate-time flow supplies such a selection.  For rigid rotation,
the generator ceases to be timelike at the speed-of-light cylinder and
$H-\Omega_{\RR}J_z$ is unbounded below on unbounded space.  For TT
rotation, coordinate-time translation is not an isometry on any
nonempty open radial region and hence supplies no conserved generator
with respect to which a stationary ground state could be selected.
These statements do not exclude the Minkowski vacuum or other
Hadamard states, canonical quantization based on an independently
chosen complex structure, rotating states defined by additional
boundary data, or stationary detector response along an individual
circular orbit.  What follows summarizes the logical conditions and
verifies these conclusions case by case in
Secs.~\ref{sec:rigid-obstruction} and \ref{sec:TT-obstruction}.

Let \(Z\) be the generator of the observer congruence on the region
\(\mathcal O\).  The stationary symmetry-selection prescription tested
here has three requirements: (a) \(Z\) is Killing; (b) \(Z\) is timelike throughout
\(\mathcal O\), with \(\mathcal O\) globally hyperbolic; and (c) the
induced one-particle generator is self-adjoint and bounded below on its
domain.  Condition (a) is what supplies a conserved, time-independent
symmetry Hamiltonian and hence a preferred stationary
positive-frequency splitting.  Condition (b) is the classical
counterpart of (c): when \(Z\) is timelike and future directed and the
matter obeys the dominant energy condition, the conserved energy
density \(T_{ab}Z^{a}n^{b}\) is nonnegative for any future-directed
timelike \(n^{b}\).  Conditions (a)--(c) are not new.  Taken together they are the hypotheses
under which the one-particle structure of a stationary linear field is
constructed, and shown to be unique, in the classic treatments of
Refs.~\cite{AshtekarMagnon1975,Kay1978,Wald1994QFTCS}, and they are the
same hypotheses under which ground and KMS states of a stationary field
are characterized in Ref.~\cite{Sanders2013}.  Condition (c) in
particular is the spectral requirement whose failure for
\(H-\Omega_{\RR}J_z\) on unbounded space has been recorded repeatedly in
the rotating literature
\cite{Vilenkin1980RotatingSystem,
LetawPfautsch1981StationaryCoordinates,
DuffyOttewill2003RotatingThermal,AmbrusWinstanley2021Rotation}.  What is
assembled here is not any one of the conditions but the comparison
between the two rotating constructions, and it is the organizing
statement of this section:
\begin{quote}
\emph{Rigid rotation satisfies} (a), \emph{but fails} (b) \emph{at}
\(r=|\Omega_{\RR}|^{-1}\) \emph{and fails} (c) \emph{on unbounded
space}, as shown in Sec.~\ref{sec:rigid-obstruction}; \emph{TT rotation
satisfies} (b) \emph{along its congruence but fails} (a), as shown in
Sec.~\ref{sec:TT-obstruction}.  \emph{The two constructions therefore
fail for complementary reasons, and the device that repairs one---a
boundary inside the speed-of-light cylinder, in the rigid case---does
nothing for the other.}
\end{quote}

\subsection{Rigid rotation}
\label{sec:rigid-obstruction}

The rigid generator \(K_{\RR}\) in Eq.~\eqref{eq:rigid-Killing} is
timelike only inside the speed-of-light cylinder.  It therefore cannot
serve as a timelike stationary evolution throughout unbounded
Minkowski spacetime.  Merely discarding the exterior does not produce
the same situation as the Rindler wedge: the open region
\(r<|\Omega_{\RR}|^{-1}\) excludes the timelike cylinder at
\(r=|\Omega_{\RR}|^{-1}\).  Causal diamonds within the open region can
contain sequences whose limit points lie on this excluded cylinder,
 and hence need not be compact as subsets of the region.  Regarded as a
 boundaryless spacetime, the open region is not globally hyperbolic.  A
 well-posed field theory on a cylinder with timelike boundary instead
 requires the boundary and its boundary conditions to be specified as
 additional data \cite{Wald1994QFTCS,Sanders2013}.  By contrast, the Rindler wedge
is itself globally hyperbolic and its relevant boundaries are null.
Quantization on spacetimes that fail to be globally hyperbolic is
possible under weaker locality assumptions, but requires additional
structure and does not by itself supply a preferred state
\cite{Kay1992Locality,Kay1997NonGlobal}.

There is also a spectral obstruction on the full unbounded space.  For
a cylindrical mode \eqref{eq:cylindrical-modes},
\(\omega=\sqrt{q^2+k^2+\mu^2}\) is the frequency
with respect to the inertial time \(T\) of Eq.~\eqref{eq:inertial-KG},
that is, the eigenvalue of \(\ii\partial_T\).  In the rigid coordinates
\eqref{eq:rigid-transformation} its phase
becomes \(\exp[-\ii(\omega-m\Omega_{\RR})\tau+\ii m\varphi]\), so its
frequency with respect to the rigid coordinate time \(\tau\), that is,
the eigenvalue of
\(\ii\partial_\tau=\ii(\partial_T+\Omega_{\RR}\partial_\Phi)\), is
\begin{equation}
 \widetilde\omega=\omega-m\Omega_{\RR}.
 \label{eq:corotating-frequency}
\end{equation}
Equivalently, \(\widetilde\omega\) is the eigenvalue of the corotating
generator \(K_{\RR}\) of Eq.~\eqref{eq:rigid-Killing}, whose quantum
counterpart is \(H-\Omega_{\RR}J_z\), with \(H\) the generator of
inertial time translations and \(J_z\) the generator of rotations about
the \(z\) axis, so that \(J_z\) has eigenvalue \(m\) on the modes
\eqref{eq:cylindrical-modes}.  In the absence of a boundary,
\(m\in\mathbb Z\) is unrestricted while,
in the cylindrical basis \eqref{eq:cylindrical-modes}, \(\omega\) is
independent of
\(m\) at fixed \(q,k\).  Hence \(H-\Omega_{\RR}J_z\) is unbounded from
below.  Sorting modes solely by the sign of
\(\widetilde\omega\) cannot define a global bosonic rotating ground
state in unbounded Minkowski spacetime
\cite{Vilenkin1980RotatingSystem,DuffyOttewill2003RotatingThermal,
AmbrusWinstanley2014RotatingStates,AmbrusWinstanley2021Rotation}.

The mirror construction of
Ref.~\cite{DuffyOttewill2003RotatingThermal} removes both problems by
changing the region and its mode spectrum.  Bounded rotating
configurations of this kind have a longer history: a detector rotating
inside a concentric circular cavity was analyzed in
Ref.~\cite{LevinPelegPeres1993}, and the conditions under which a
rotating detector does or does not respond were derived for a range of
bounded spacetimes in
Ref.~\cite{DaviesDrayManogue1996RotatingVacuum}; see also the null-response
analysis of Ref.~\cite{Nicolaevici2001NullResponse}.

The combination \(\omega-m\Omega_{\RR}\) is also the kinematic
combination that appears in rotational superradiance
\cite{Zeldovich1971,BekensteinSchiffer1998}.  In both settings a mode
may have positive inertial frequency and positive Klein--Gordon norm
but negative energy relative to the corotating generator.  This analogy
does not turn the detector calculation of Sec.~\ref{sec:detector} into
a scattering or
ergoregion-instability calculation: a genuine ergoregion can trap
negative Killing energy \cite{Friedman1978}, and negative corotating
frequency alone is not sufficient to establish a vacuum instability
\cite{MataczDaviesOttewill1993}.

A cylindrical mirror defines a different physical problem by replacing
unbounded Minkowski space with a timelike-boundary value problem.  This
replacement, and its consequences for the response of a rotating
detector, are precisely what
Refs.~\cite{LevinPelegPeres1993,DaviesDrayManogue1996RotatingVacuum,
Nicolaevici2001NullResponse,DuffyOttewill2003RotatingThermal} analyze;
we recall the spectral mechanism here only because it isolates what the
unbounded problem lacks.
For a Dirichlet mirror at \(r=R_0<|\Omega_{\RR}|^{-1}\), with
\(m\in\mathbb Z\), the radial momenta are quantized as
\begin{equation}
 q_{mn}=\frac{j_{|m|,n}}{R_0},
 \label{eq:mirror-momenta}
\end{equation}
where \(j_{\nu,n}\) is the \(n\)th positive zero of \(J_\nu\).  The
positivity of the corotating frequency then follows in three steps.
First, \(j_{|m|,n}>|m|\) for every \(n\geq1\), so that
\(q_{mn}>|m|/R_0\).  Second, the corotating frequency is built from
\(\omega\) rather than from \(q\), and since
\(\omega_{mn}=\sqrt{q_{mn}^2+k^2+\mu^2}\geq q_{mn}\), the same lower
bound propagates to \(\omega_{mn}>|m|/R_0\).  Third---and this is where
the placement of the mirror inside the speed-of-light cylinder does its
work---the hypothesis \(R_0<|\Omega_{\RR}|^{-1}\) means precisely that
\(1/R_0>|\Omega_{\RR}|\), so that \(|m|/R_0\geq|m\Omega_{\RR}|\), with
equality only at \(m=0\).  Chaining the three,
\begin{equation}
 \omega_{mn}>\frac{|m|}{R_0}\geq m\Omega_{\RR},
 \label{eq:mirror-positivity}
\end{equation}
including the modes that could otherwise have negative corotating
frequency, namely those with \(m\Omega_{\RR}>0\).  The first inequality
is strict for every \(m\), so every corotating frequency
\(\widetilde\omega_{mn}=\omega_{mn}-m\Omega_{\RR}\) is strictly
positive; at \(m=0\) the second inequality is saturated and the
statement reduces to \(\widetilde\omega_{0n}=\omega_{0n}>0\).  Once
the boundary condition and self-adjoint dynamics are specified, a
ground or KMS construction can be posed on this new region: the
boundary lies where \(K_{\RR}\) is timelike and the allowed spectrum is
positive \cite{Sanders2013}.  This construction does not
solve or modify the response of a detector rotating in unbounded
Minkowski spacetime; it defines another setup
\cite{DaviesDrayManogue1996RotatingVacuum,
Nicolaevici2001NullResponse,DuffyOttewill2003RotatingThermal}.

\subsection{Trocheris--Takeno rotation}
\label{sec:TT-obstruction}

The vector field tangent to every curve of constant TT spatial
coordinates is the TT coordinate-time vector
\begin{equation}
 \KTT=\partial_t
 =C(r)\partial_T+\frac{S(r)}r\partial_\Phi,
 \label{eq:TT-flow}
\end{equation}
with \(C\) and \(S\) the hyperbolic functions of the radial rapidity
defined in Eq.~\eqref{eq:TT-CS}.  We write it as \(\KTT\) rather than as a Killing field \(K\), because it
is not one: it has unit norm, but it fails to be Killing.  Two
components of the Lie derivative along \(\KTT\) of the Minkowski metric
in inertial cylindrical coordinates, Eq.~\eqref{eq:inertial-metric},
already prove the point:
\begin{equation}
 (\mathcal L_{\KTT}g)_{rT}
 =\Omega_{\TT}S,\qquad
 (\mathcal L_{\KTT}g)_{r\Phi}=S-\eta C.
 \label{eq:TT-Lie}
\end{equation}
They are nonzero at generic \(r\).  Consequently, translation in \(t\)
is not an isometry, and the coefficient multiplying \(t\) in the phase
of a mode is not constant, so it cannot be a conserved frequency.

The failure to be Killing has a transparent origin, and it is worth
isolating, because the same identity explains why individual TT orbits are stationary while the whole congruence does not define a Killing field.  The Lie derivative is
additive in its generator, and homogeneous only under multiplication by
constants:
\begin{equation}
 \mathcal L_{X+Y}=\mathcal L_X+\mathcal L_Y,
 \qquad
 \mathcal L_{cX}=c\,\mathcal L_X
 \quad (c\ \text{constant}).
 \label{eq:Lie-linearity}
\end{equation}
For a \emph{function} \(f\) the homogeneity fails, and acting on the
metric one has instead
\begin{equation}
 (\mathcal L_{fX}g)_{ab}
 =f\,(\mathcal L_{X}g)_{ab}
 +X_a\nabla_b f+X_b\nabla_a f.
 \label{eq:Lie-function}
\end{equation}
The coefficients \(C(r)\) and \(S(r)/r\) in Eq.~\eqref{eq:TT-flow}
depend on \(r\), so the inhomogeneous terms in
Eq.~\eqref{eq:Lie-function} survive; they are precisely the radial
components displayed in Eq.~\eqref{eq:TT-Lie}.

Evolution generated by this same vector field \(\partial_t\) also does
not define a global Cauchy slicing; that is, it does not slice the spacetime with Cauchy surfaces.  Besides the identification
\eqref{eq:TT-identification}, one may evaluate the norm of the gradient
of the covering coordinate \(t=CT-rS\Phi\) of
Eq.~\eqref{eq:TT-forward}.  Working in the inertial cylindrical
coordinates \((T,r,\Phi,z)\) and using the metric
\eqref{eq:inertial-metric}, so that
\(g^{ab}(\nabla_a t)(\nabla_b t)
=(\partial_T t)^2-(\partial_r t)^2-r^{-2}(\partial_\Phi t)^2\), one
finds
\begin{equation}
 g^{ab}(\nabla_a t)(\nabla_b t)
 =1-\left[
 \Omega_{\TT}ST-(S+\eta C)\Phi
 \right]^2.
 \label{eq:TT-time-gradient}
\end{equation}

As specified in Sec.~\ref{sec:TT-map},
\(-\infty<\Phi<\infty\) on the angular covering space on which
Eq.~\eqref{eq:TT-forward} defines \(t\).  The bracket in
Eq.~\eqref{eq:TT-time-gradient} then grows without bound, so
Eq.~\eqref{eq:TT-time-gradient} becomes negative for sufficiently large
\(\Phi\).  Where \(\nabla_a t\) is spacelike
the level set \(t=\mathrm{const}\) is a timelike hypersurface, so those
surfaces do not remain spacelike.  A canonical
quantization prescription can still be built on an inertial
Cauchy surface, for instance \(T=\mathrm{const}\), and rewritten in TT
coordinates; what is not available is a global TT-stationary vacuum
prescription.

The restriction of \(\KTT\) to any one orbit is, however,
proportional to a Killing field, and this is what makes the response
function of Sec.~\ref{sec:response-functional} stationary on that orbit.
Indeed, at \(r=R\), let
\begin{equation}
 \Omega_R\equiv\Omega_{\mathrm{phys}}^{\TT}(R),\qquad
 \Xi_R=\partial_T+\Omega_R\partial_\Phi.
 \label{eq:one-orbit-Killing}
\end{equation}
For each fixed \(R\), \(\Omega_R\) is a constant, so the additivity and
constant-homogeneity in Eq.~\eqref{eq:Lie-linearity} apply, and since
\(\partial_T\) and \(\partial_\Phi\) are Killing fields of Minkowski
spacetime,
\begin{equation}
 \mathcal L_{\Xi_R}g=
 \mathcal L_{\partial_T}g+\Omega_R\mathcal L_{\partial_\Phi}g=0.
 \label{eq:Xi-Killing}
\end{equation}
The vector \(\KTT\) from Eq.~\eqref{eq:TT-flow}, restricted to that
orbit, is the normalized
form of \(\Xi_R\),
\begin{equation}
 \left.\KTT\right|_{r=R}
 =\cosh(\Omega_{\TT}R)\,\Xi_R.
 \label{eq:KTT-orbit}
\end{equation}
Equations~\eqref{eq:Xi-Killing} and \eqref{eq:TT-Lie} illustrate the
difference between the two cases in
Eqs.~\eqref{eq:Lie-linearity} and \eqref{eq:Lie-function}: at fixed
\(R\), \(\Omega_R\) is a number and can be pulled through the Lie
derivative, whereas the full-congruence coefficients are functions of
\(r\) and generate the additional derivative terms.
Each TT circle is therefore an orbit of a Killing field, but a different
field \(\Xi_R\) is needed at each radius, and no single Killing field
generates the congruence.  This is sufficient for
stationarity of the Minkowski-vacuum positive-frequency two-point
function restricted to one such
worldline, and it is not sufficient for a stationary
quantization of the full TT
congruence.  This orbitwise statement is an instance of the
classification of stationary worldlines as timelike Killing orbits
\cite{Letaw1981StationaryWorldLines,Bunney2024}; the seminal
rotating-coordinate analyses already emphasize that a stationary
response on one observer orbit and a vacuum selected by a global
time-translation symmetry are distinct notions
\cite{LetawPfautsch1980RotatingCoordinates,
LetawPfautsch1981StationaryCoordinates}.

\section{One Minkowski state  in three coordinate systems}
\label{sec:covariance}

The purpose of this section is to separate a passive coordinate
rewriting from a change of complex structure.  For modes explicitly
defined as coordinate representatives of a fixed inertial basis, the
vanishing-beta result is a consistency statement, not a
dynamical discovery. For a covariant theory, only a change of coordinates should not modify any of its contents, which in this case are the Minkowski vacuum state and its related two-point function. The substantive checks are whether the displayed
functions are globally admissible and correctly normalized, and whether
their complete mode sums reproduce the chosen state's two-point
function.  Consider a coordinate map, written in the sense fixed in
Sec.~\ref{sec:conventions},
\[
 F:U\longrightarrow\mathcal M
\]
from coordinates \(x\in U\) to the corresponding Minkowski event
\(X=F(x)\).  Here \(U\) is an open subset of \(\mathbb R^{4}\)---a
coordinate domain, not a spacetime in its own right---and \(F\) is a
smooth map with smooth inverse onto its image
\(F(U)\subseteq\mathcal M\), which is where the metric, the field, and
the state actually live.  The two cases used below differ in what
\(F(U)\) is.  The rigid map \eqref{eq:rigid-transformation} covers all
of \(\mathcal M\), up to the degeneracy that cylindrical coordinates
already have on the axis, and in particular it is not obstructed by the
speed-of-light cylinder.  The TT map \eqref{eq:TT-inverse} is defined
for \(r>0\) on the angular cover, so it is a covering rather than a
global chart, and the identification \eqref{eq:TT-identification} must
be carried along with it.  Since the field is a scalar, its coordinate expression is
\(\phi_F(x)=\phi(F(x))\).  The Wightman function in the new coordinates
is therefore
\begin{equation}
 \Wight_F(x,x')
 =\Wight_{\M}\!\left(F(x),F(x')\right).
 \label{eq:W-coordinate-change}
\end{equation}

Both spacetime arguments are transformed.  Equation
\eqref{eq:W-coordinate-change} changes only the coordinate expression
of the Minkowski-vacuum correlation function; it does not select a new
state.

The reason Eq.~\eqref{eq:W-coordinate-change} takes this particularly
simple form is the spin of the field, not any property of the state.  A
scalar carries no coordinate indices, so its value at a spacetime point is a
number and its coordinate expression is the composition
\(\phi_F=\phi\circ F\).  Consequently \(\Wight\) is a biscalar: it is a
number attached to an ordered pair of events, and a passive coordinate
change can do nothing to it but relabel its two arguments.

Let \(v_\lambda(X)\) be an inertial mode, that is, a positive-frequency
solution of Eq.~\eqref{eq:inertial-KG} normalized as in
Eq.~\eqref{eq:cylindrical-normalization}.  The same solution written in
the new coordinates is
\begin{equation}
 u_\lambda(x)=v_\lambda(F(x)).
 \label{eq:transformed-mode}
\end{equation}
Before a mode comparison is meaningful, the functions
\(u_\lambda\) must be admissible solutions on the same physical
spacetime: they must solve the transformed wave equation and respect its
global identifications.  Local covariance guarantees the first property
for Eq.~\eqref{eq:transformed-mode}, but invariance under any
angular identification $\Phi \sim \Phi+2\pi$ must still be checked.  Sections~\ref{sec:rigid-field} and
\ref{sec:TT-field} verify both properties explicitly.

To compare the two coordinate representatives without identifying
functions written on different coordinate covers or charts, let $\widehat u_\lambda(X)$ be the transformed mode
back in inertial coordinates:
\begin{equation}
 \widehat u_\lambda(X)
 \equiv u_\lambda(F^{-1}(X))
 =v_\lambda(X).
 \label{eq:mode-pushback-identity}
\end{equation}

\noindent Equation~\eqref{eq:mode-pushback-identity} is an identity following
from the definition \eqref{eq:transformed-mode}; it does not use the
field dynamics or determine a new positive-frequency splitting.

Since \(\{v_\lambda,v^{*}_\lambda\}\) is a basis of the inertial
solution space, expand
\begin{equation}
 \widehat u_\lambda=\sum_{\lambda'}
 \left(
 \alpha_{\lambda\lambda'}v_{\lambda'}
 +\beta_{\lambda\lambda'}v^{*}_{\lambda'}
 \right),
 \label{eq:bogoliubov-expansion}
\end{equation}
which defines the Bogoliubov coefficients
\(\alpha_{\lambda\lambda'}\) and \(\beta_{\lambda\lambda'}\).  With the
antilinear-first convention of Eq.~\eqref{eq:KG-product}, orthonormality
gives
\begin{equation}
 \alpha_{\lambda\lambda'}=\KG{v_{\lambda'}}{\widehat u_\lambda},
 \qquad
 \beta_{\lambda\lambda'}=-\KG{v^{*}_{\lambda'}}{\widehat u_\lambda}.
 \label{eq:bogoliubov-overlaps}
\end{equation}
A nonzero \(\beta\) mixes annihilation with creation operators and hence
changes the positive-frequency subspace.  In the present case,
Eq.~\eqref{eq:mode-pushback-identity} and uniqueness of the basis
expansion give directly
\begin{equation}
 \alpha_{\lambda\lambda'}=\delta_{\lambda\lambda'},
 \qquad
 \beta_{\lambda\lambda'}=0.
 \label{eq:coordinate-beta}
\end{equation}
Thus Eq.~\eqref{eq:coordinate-beta} applies specifically to the
coordinate-rewritten modes \eqref{eq:transformed-mode}, compared with
their parent inertial basis.  It does not assert that every admissible
mode basis expressed in rotating coordinates must have vanishing beta
coefficients.

The Klein--Gordon product is coordinate invariant, so the overlaps in
Eq.~\eqref{eq:bogoliubov-overlaps} may be evaluated in either coordinate representation, but
they must refer to the same geometric Cauchy surface.  We use
\(T=\mathrm{const}\) for convenience.  In TT coordinates that same
surface is \(Ct+rS\theta=\mathrm{const}\); it must not be replaced by
\(t=\mathrm{const}\), because the latter is not a global Cauchy surface
by Eq.~\eqref{eq:TT-time-gradient} and the identification
\eqref{eq:TT-identification}.

The content beyond this identity lies in the checks
carried out below: the rigid and TT expressions must solve the exact
transformed equations, satisfy the angular identifications, be
normalized on a common geometric Cauchy surface, and reconstruct the
complete Minkowski Wightman function.  A genuinely different vacuum
would instead require an independently chosen admissible complex
structure.  The present beta-zero statement neither constructs nor
excludes such an additional choice; Sec.~\ref{sec:no-go} addresses only
whether the corresponding rotating time flows select one.

\subsection{Rigid-coordinate modes and Wightman function}
\label{sec:rigid-field}

Using Eq.~\eqref{eq:rigid-transformation}, the Klein--Gordon equation is
\begin{equation}
 \left[
 (\partial_\tau-\Omega_{\RR}\partial_\varphi)^2
 -\frac1r\partial_r(r\partial_r)
 -\frac1{r^2}\partial_\varphi^2-\partial_z^2+\mu^2
 \right]\phi=0.
 \label{eq:rigid-KG}
\end{equation}
Equation~\eqref{eq:rigid-KG} is the wave operator of the geometry and
not merely the inertial operator carried along by the chain rule, as the
covariant expression
\begin{equation}
 \Box\phi=\frac1{\sqrt{-g}}\,
 \partial_\mu\!\left(\sqrt{-g}\,g^{\mu\nu}\partial_\nu\phi\right)
 \label{eq:box-covariant}
\end{equation}
shows when it is evaluated with the rigid metric
\eqref{eq:rigid-metric} alone.  The \((\tau,\varphi)\) block of
Eq.~\eqref{eq:rigid-metric} has determinant \(-r^{2}\) by
Eq.~\eqref{eq:rigid-block-determinant}, and \(g_{rr}=g_{zz}=-1\), so
\(\det g_{\RR}=-r^{2}\) and \(\sqrt{-g}=r\), the same measure factor as
in the inertial cylindrical chart.  Inverting the same block gives
\begin{equation}
 g^{\tau\tau}_{\RR}=1,\quad
 g^{\tau\varphi}_{\RR}=-\Omega_{\RR},\quad
 g^{\varphi\varphi}_{\RR}=\Omega_{\RR}^{2}-\frac1{r^{2}},
 \label{eq:rigid-inverse-metric}
\end{equation}
together with \(g^{rr}_{\RR}=g^{zz}_{\RR}=-1\) and no further mixing.
Substituting Eq.~\eqref{eq:rigid-inverse-metric} into
Eq.~\eqref{eq:box-covariant}, and using that \(r\) is the only
coordinate on which the components depend, gives
\begin{align}
 \Box_{\RR}\phi
 ={}&\partial_\tau\!\left(
 \partial_\tau\phi-\Omega_{\RR}\partial_\varphi\phi\right)
 -\frac1r\partial_r\!\left(r\partial_r\phi\right)
 \nonumber\\
 &+\partial_\varphi\!\left[
 -\Omega_{\RR}\partial_\tau\phi
 +\left(\Omega_{\RR}^{2}-\frac1{r^{2}}\right)
 \partial_\varphi\phi\right]
 -\partial_z^{2}\phi
 \nonumber\\
 ={}&(\partial_\tau-\Omega_{\RR}\partial_\varphi)^{2}\phi
 -\frac1r\partial_r(r\partial_r\phi)
 -\frac1{r^{2}}\partial_\varphi^{2}\phi
 -\partial_z^{2}\phi,
 \label{eq:rigid-box-covariant}
\end{align}
which is Eq.~\eqref{eq:rigid-KG}.  The corresponding covariant
calculation for the TT chart, where the metric functions depend on
\(t\) and \(\theta\) as well and the regrouping is less immediate, is
carried out in Appendix~\ref{app:TT-operator}.

An orthonormal set of positive-inertial-frequency solutions of
Eq.~\eqref{eq:rigid-KG} is
\begin{equation}
 u^{\RR}_{qmk}(y)=
 \frac1{2\pi}\sqrt{\frac{q}{2\omega}}\,
 J_m(qr)\e^{\ii kz+\ii m\varphi}
 \e^{-\ii(\omega-m\Omega_{\RR})\tau}.
 \label{eq:rigid-modes}
\end{equation}
Even when \(\omega-m\Omega_{\RR}<0\), the mode has positive
Klein--Gordon norm.  This point is central to the argument, so it is
worth making explicit.  The norm \eqref{eq:KG-product} is evaluated on
the inertial Cauchy surface \(T=\mathrm{const}\), whose future-directed
unit normal is \(n^{a}=(\partial_T)^{a}\); the integrand therefore
involves \(\ii(f^{*}\partial_T f-f\partial_T f^{*})\), and its sign is
fixed by the eigenvalue of \(\ii\partial_T\), namely the inertial
frequency \(\omega>0\).  The corotating frequency
\(\widetilde\omega=\omega-m\Omega_{\RR}\) is the eigenvalue of a
\emph{different} operator, \(\ii\partial_\tau\), and it does not enter
the Klein--Gordon norm \(\KG{\cdot}{\cdot}\) of
Eq.~\eqref{eq:KG-product} at all.  Sign of norm and sign of corotating frequency are
thus independent, and
\begin{equation}
 \alpha^{\RR}_{\lambda\lambda'}=\delta_{\lambda\lambda'},
 \qquad \beta^{\RR}_{\lambda\lambda'}=0,
 \label{eq:rigid-beta}
\end{equation}
as required by the coordinate identity
Eq.~\eqref{eq:coordinate-beta}.  The useful check here is that a mode
with negative corotating frequency retains positive inertial frequency
and positive Klein--Gordon norm.  The same distinction
reappears in Sec.~\ref{sec:circular-excitation}, where a mode of
negative corotating frequency and positive norm is what allows the
detector to be excited.

The admissibility and common-Cauchy-surface requirements of
Sec.~\ref{sec:covariance} hold here.  The rigid chart shares the inertial time coordinate,
\(T=\tau\) in Eq.~\eqref{eq:rigid-transformation}, so surfaces of
constant \(\tau\) are the inertial Cauchy surfaces themselves; and
because \(\varphi\) enters Eq.~\eqref{eq:rigid-modes} only through
\(\e^{\ii m\varphi}\) with \(m\in\mathbb Z\), the modes are
single-valued under \(\varphi\sim\varphi+2\pi\).  Direct substitution
verifies that they solve Eq.~\eqref{eq:rigid-KG}.

The Wightman function calculated from these cylindrical modes is
\begin{align}
 \Wight_{\RR,\mu}(y,y')
 &=\sum_{m=-\infty}^{\infty}
 \int_0^\infty\dd q\int_{-\infty}^{\infty}\dd k\,
 \frac{q}{8\pi^2\omega}
 \nonumber\\
 &\quad\times J_m(qr)J_m(qr')
 \e^{\ii m(\Delta\varphi+\Omega_{\RR}\Delta\tau)}
 \nonumber\\
 &\quad\times
 \e^{\ii k\Delta z-\ii\omega(\Delta\tau-\ii\epsilon)}.
 \label{eq:rigid-W-sum}
\end{align}
This has the same Bessel sum as Eq.~\eqref{eq:inertial-mode-sum}, with
\[
 \Delta T=\Delta\tau,\qquad
 \Delta\Phi=\Delta\varphi+\Omega_{\RR}\Delta\tau.
\]
Applying Graf's identity exactly as in Sec.~\ref{sec:graf} gives, for a
massless field,
\begin{equation}
 \Wight_{\RR,0}(y,y')
 =\frac{1}{4\pi^2\rho_{\RR,\epsilon}(y,y')},
 \label{eq:rigid-W}
\end{equation}
where
\begin{align}
 \rho_{\RR,\epsilon}
 ={}&-(\Delta\tau-\ii\epsilon)^2+r^2+r'^2
 -2rr'\cos(\Delta\varphi+\Omega_{\RR}\Delta\tau)
 \nonumber\\
 &+(\Delta z)^2.
 \label{eq:rigid-rho}
\end{align}
The expression for a massive field follows by replacing
\((4\pi^2\rho)^{-1}\) with the right-hand side of
Eq.~\eqref{eq:massive-W}.  Equations
\eqref{eq:rigid-W}--\eqref{eq:rigid-rho} can be written
as
\begin{equation}
 \Wight_{\RR}(y,y')
 =\Wight_{\M}\!\left(F_{\RR}(y),F_{\RR}(y')\right).
 \label{eq:rigid-W-transformation}
\end{equation}
The beta-zero statement was already fixed by the coordinate
identity.  The mode sum provides a separate and stronger consistency
check: the complete normalized rigid basis reconstructs the Minkowski
Wightman function for arbitrary pairs of spacetime points.

\subsection{TT modes and Wightman function}
\label{sec:TT-field}

Because the TT boost parameter depends on \(r\), the radial part of the
wave operator contains mixed derivatives.  A compact exact expression
uses the metric functions in Eq.~\eqref{eq:TT-metric-functions} and the
modified radial derivative
\begin{equation}
 \mathfrak D_r=\partial_r-\mathfrak A\,\partial_t
 +\frac{\mathfrak B}{r^2}\partial_\theta.
 \label{eq:TT-Dr}
\end{equation}
The exact TT Klein--Gordon equation is
\begin{equation}
 \left[
 \partial_t^2-\frac1r\mathfrak D_r(r\mathfrak D_r)
 -\frac1{r^2}\partial_\theta^2-\partial_z^2+\mu^2
 \right]\phi=0.
 \label{eq:TT-KG}
\end{equation}
We define \(\Box_{\TT}\) as the differential operator in square brackets
in Eq.~\eqref{eq:TT-KG} without the mass term.
Appendix~\ref{app:TT-operator} presents the derivation of
Eq.~\eqref{eq:TT-KG}.

The TT Klein--Gordon equation has the following orthonormal set of
positive-inertial-frequency solutions:
\begin{equation}
 \begin{aligned}
 u^{\TT}_{qmk}(x)
 &=\frac1{2\pi}\sqrt{\frac{q}{2\omega}}\,
 J_m(qr)\e^{\ii kz}\\
 &\quad\times
 \exp\!\left[
 \ii(mC-\omega rS)\theta
 -\ii\left(\omega C-\frac{mS}{r}\right)t
 \right].
 \end{aligned}
 \label{eq:TT-modes}
\end{equation}
The labels run over the same ranges as in
Eq.~\eqref{eq:cylindrical-modes}, namely \(q>0\), \(m\in\mathbb Z\) and
\(k\in\mathbb R\), and \(\omega=\sqrt{q^{2}+k^{2}+\mu^{2}}\) is the same
inertial frequency as there; \(C\) and \(S\) are the functions
\eqref{eq:TT-CS} evaluated at the radius \(r\) of the argument.  These are not obtained by separating Eq.~\eqref{eq:TT-KG}; they are the
inertial cylindrical modes \eqref{eq:cylindrical-modes} composed with
the map \(F_{\TT}\) of Eq.~\eqref{eq:TT-inverse}, in accordance with
Eq.~\eqref{eq:transformed-mode}.  Direct substitution into
Eq.~\eqref{eq:TT-KG} then confirms
\begin{equation}
 (\Box_{\TT}+\mu^2)u^{\TT}_{qmk}=0.
 \label{eq:TT-mode-check}
\end{equation}
This verifies the local field-equation requirement stated in
Sec.~\ref{sec:covariance}.  Global admissibility also requires the
solutions to be single-valued under the
identification \eqref{eq:TT-identification}.  Going once around, their
phase changes by
\begin{align}
 \Delta\chi
 &=2\pi\left[
 C(mC-\omega rS)
 +rS\left(\omega C-\frac{mS}{r}\right)
 \right]\nonumber\\
 &=2\pi m,
 \label{eq:TT-single-valued}
\end{align}
so that \(\e^{\ii\Delta\chi}=1\) for \(m\in\mathbb Z\).  The mode
is therefore invariant under the helical identification and descends to
a single-valued mode on physical Minkowski spacetime, exactly as the
cylindrical modes \eqref{eq:cylindrical-modes} are under
\(\Phi\sim\Phi+2\pi\); the \(r\)-dependent coefficients in the exponent
of Eq.~\eqref{eq:TT-modes} conspire to cancel.

The physical spacetime is recovered as the quotient of the 
TT covering space by the discrete isometry group $\Gamma \cong \mathbb{Z}$ 
generated by the helical shift \eqref{eq:TT-identification}. More generally, 
quantum fields on a quotient manifold $\widetilde{\mathcal{M}}/\Gamma$ may 
transform under any one-dimensional unitary representation of the identification 
group, giving rise to distinct automorphic or twisted sectors 
\cite{BanachDowker1979Math,BanachDowker1979Stress,Sushkov1997}. Such choices 
constitute additional global boundary data: while sharing identical local field 
equations and short-distance singularities in any single chart, different sectors 
introduce relative signs or phase factors into the image sums that build the 
Wightman function, leading to genuinely different transition rates for an 
Unruh--DeWitt detector along the same trajectory \cite{Langlois2006}. For the 
standard real scalar field in Minkowski spacetime, one strictly requires the 
trivial (untwisted) representation, where the field is single-valued under 
$\Gamma$. This is precisely what the integer phase matching $\mathrm{e}^{\mathrm{i}\Delta\chi} = 1$ 
in Eq.~\eqref{eq:TT-single-valued} guarantees; twisted TT sectors define alternative 
field theories and are not considered here.

To verify the
normalization consistently with Sec.~\ref{sec:covariance}, we evaluate
the Klein--Gordon product on the same Cauchy surface of \(\mathcal M\)
used in the inertial calculation,
\(T=\mathrm{const}\), whose TT coordinate equation is
\(Ct+rS\theta=\mathrm{const}\).  This gives
\begin{equation}
 \begin{aligned}
 \KG{u^{\TT}_{qmk}}{u^{\TT}_{q'm'k'}}
 &=\delta(q-q')\delta_{mm'}\delta(k-k'),\\
 \KG{u^{\TT}_{qmk}}{u^{\TT *}_{q'm'k'}}&=0.
 \end{aligned}
 \label{eq:TT-normalization}
\end{equation}
These relations verify normalization and positive/negative-norm
orthogonality on a valid Cauchy surface.  They support the admissibility
of the TT mode basis; the value \(\beta^{\TT}=0\) relative to the parent
inertial basis has already followed from
Eq.~\eqref{eq:mode-pushback-identity}.

The details of this computation are given in
Appendix~\ref{app:TT-KG-normalization}.  Normalization on \(t=\mathrm{const}\)
would not establish a global quantization, because those surfaces are
not Cauchy surfaces.  There are two independent reasons.  First, by Eq.~\eqref{eq:TT-time-gradient} the gradient
\(\nabla_a t\) becomes spacelike at sufficiently large
\(\Phi\), so the level sets \(t=\mathrm{const}\) become timelike
hypersurfaces there; a Cauchy surface must be
achronal,\footnote{A set is \emph{achronal} when no two of its points
can be joined by a timelike curve, so that no point of the set lies in
the chronological future of another
\cite{Wald1994QFTCS,HawkingEllis1973}.  A timelike hypersurface always
contains such pairs.} and a timelike
hypersurface is not.  Second, by Eq.~\eqref{eq:TT-identification},
the covering coordinate \(t\) does not descend to a single-valued
function on physical Minkowski spacetime, so its
level sets are not well defined globally to begin with.

Let \(x=(t,r,\theta,z)\) and
\(x'=(t',r',\theta',z')\) be two arbitrary TT covering-coordinate
representatives, not necessarily points on the same circular orbit.  Their
inertial time and angle coordinates are
\begin{align}
 T_x&=Ct+rS\theta,&
 \Phi_x&=C\theta+\frac Sr t,\nonumber\\
 T_{x'}&=C't'+r'S'\theta',&
 \Phi_{x'}&=C'\theta'+\frac{S'}{r'}t'.
 \label{eq:TT-images}
\end{align}
Multiplying \(u^{\TT}_{qmk}(x)\) by
\(u^{\TT *}_{qmk}(x')\) and re-expressing the whole exponent of
Eq.~\eqref{eq:TT-modes} in terms of
Eq.~\eqref{eq:TT-images} gives
\begin{align}
 \Wight_{\TT,\mu}(x,x')
 &=\sum_{m=-\infty}^{\infty}
 \int_0^\infty\dd q\int_{-\infty}^{\infty}\dd k\,
 \frac{q}{8\pi^2\omega}
 \nonumber\\
 &\quad\times J_m(qr)J_m(qr')
 \e^{\ii m(\Phi_x-\Phi_{x'})}
 \nonumber\\
 &\quad\times
 \e^{\ii k(z-z')-\ii\omega(T_x-T_{x'}-\ii\epsilon)}.
 \label{eq:TT-W-sum}
\end{align}
Appendix~\ref{app:TT-W-sum} shows the phase collection and Bessel
summation step by step.  For the massless field the result is
\begin{equation}
 \Wight_{\TT,0}(x,x')
 =\frac1{4\pi^2\rho_{\TT,\epsilon}(x,x')},
 \label{eq:TT-W}
\end{equation}
with
\begin{align}
 \rho_{\TT,\epsilon}
 ={}&-(T_x-T_{x'}-\ii\epsilon)^2
 +r^2+r'^2\nonumber\\
 &-2rr'\cos(\Phi_x-\Phi_{x'})
 +(z-z')^2.
 \label{eq:TT-rho}
\end{align}
The result can be written directly as
\begin{equation}
 \Wight_{\TT}(x,x')
 =\Wight_{\M}\!\left(F_{\TT}(x),F_{\TT}(x')\right).
 \label{eq:TT-W-transformation}
\end{equation}
This equality holds for arbitrary pairs of representatives on the
angular cover and is invariant under the helical identification in each
argument; it does not require restricting the two points to one
detector trajectory.  We therefore conclude that the TT expression is
the Minkowski-vacuum two-point function written in TT coordinates,
not a distinct vacuum defined only along a chosen circular orbit.

A remark on orientation is in order here, since mode functions of the
form \eqref{eq:TT-modes} but with the opposite signs in the exponent,
\begin{align}
 u^{\mathrm{opp}}_{qmk}
 &\propto J_m(qr)\e^{\ii kz}\nonumber\\
 &\quad\times
 \exp\!\left[
 \ii(mC+\omega rS)\theta
 -\ii\left(\omega C+\frac{mS}{r}\right)t
 \right],
 \label{eq:opposite-mode}
\end{align}
appear in earlier TT-based constructions
\cite{DeLorenciSvaiter1999Foundations,
DeLorenciEtAl2000RotatingDetector,DeLorenciEtAl2001Erratum,
DePaolaSvaiter2001Bucket}.  Read consistently, these correspond to the
opposite TT orientation in the sense of Sec.~\ref{sec:TT-map},
\[
 T=Ct-rS\theta,\qquad
 \Phi=C\theta-\frac Sr t,
\]
that is, to \(\Omega_{\TT}\to-\Omega_{\TT}\) in
Eq.~\eqref{eq:TT-inverse}.  In that reading
\begin{equation}
 u^{\mathrm{opp}}_{qmk}(x)
 =v_{qmk}\!\left(
 F_{\TT}|_{\Omega_{\TT}\to-\Omega_{\TT}}(x)\right),
 \label{eq:opposite-orientation}
\end{equation}
and the argument of Sec.~\ref{sec:covariance} applies again: these
are inertial modes written in the oppositely oriented TT covering coordinates.
Their beta coefficients relative to that inertial basis therefore
vanish by identity, while normalization and reconstruction of the
Minkowski Wightman function remain the relevant consistency checks.

\subsection{Direct rigid--TT equivalence}
\label{sec:direct-equivalence}

Let \(H=F_{\TT}^{-1}\circ F_{\RR}\), which maps rigid coordinates to TT
coordinates for the same event.  Substituting
\(T=\tau\), \(\Phi=\varphi+\Omega_{\RR}\tau\) into
Eq.~\eqref{eq:TT-forward} gives
\begin{equation}
 \begin{split}
 t&=(C-r\Omega_{\RR}S)\tau-rS\varphi,\\
 \theta&=C\varphi+
 \left(C\Omega_{\RR}-\frac Sr\right)\tau.
 \end{split}
 \label{eq:transition-map}
\end{equation}
Inserting Eq.~\eqref{eq:transition-map} into
Eq.~\eqref{eq:TT-inverse} verifies
\[
 T_{\TT}(H(y))=\tau,\qquad
 \Phi_{\TT}(H(y))=\varphi+\Omega_{\RR}\tau.
\]
The exponent of Eq.~\eqref{eq:TT-modes},
\[
 \ii(mC-\omega rS)\theta
 -\ii\left(\omega C-\frac{mS}{r}\right)t,
\]
then reduces to
\[
 \ii m\varphi-\ii(\omega-m\Omega_{\RR})\tau,
\]
so
\begin{equation}
 u^{\TT}_{qmk}(H(y))=u^{\RR}_{qmk}(y).
 \label{eq:direct-mode-equivalence}
\end{equation}
Because the transformed inertial coordinates agree for each point,
the invariant separations \(\rho_\epsilon\) of
Eq.~\eqref{eq:rho-inertial}, carrying the \(\ii\epsilon\) prescription
that fixes the positive-frequency boundary value, also agree:
\begin{equation}
 \rho_{\TT,\epsilon}(H(y),H(y'))
 =\rho_{\RR,\epsilon}(y,y'),
 \label{eq:direct-rho-equivalence}
\end{equation}
which proves
\begin{equation}
 \Wight_{\TT}(H(y),H(y'))
 =\Wight_{\RR}(y,y')
 =\Wight_{\M}(F_{\RR}(y),F_{\RR}(y')).
 \label{eq:direct-W-equivalence}
\end{equation}
Equations~\eqref{eq:direct-mode-equivalence} and
\eqref{eq:direct-W-equivalence} are a direct verification of the conditions \eqref{eq:mode-pushback-identity} and \eqref{eq:W-coordinate-change} respectively. Therefore, the consistency of the quantization in these two coordinate systems is satisfied.

\section{Circular-detector response}
\label{sec:detector}

\subsection{Response functional and stationary rate}
\label{sec:response-functional}

Let \(x_{\mathrm D}(s)\) be an arbitrary timelike detector trajectory
parametrized by proper time \(s\); it need not yet be circular.  For a
two-level detector with energy gap \(E\) and smooth compactly supported
switching function \(\chi\), the leading response---that is, the
response function at second order in the detector--field coupling---is
\cite{Unruh1976BlackHole,BirrellDavies1982QFCS,Takagi1986VacuumNoise,
CrispinoHiguchiMatsas2008Unruh,LoukoSatz2006Clicks,Satz2007}
\begin{align}
 \mathcal F(E)
 &=
 \int_{-\infty}^{\infty}\dd s
 \int_{-\infty}^{\infty}\dd s'\,
 \chi(s)\chi(s')\nonumber\\
 &\quad\times
 \e^{-\ii E(s-s')}
 \Wight_{\M}(x_{\mathrm D}(s),x_{\mathrm D}(s')).
 \label{eq:response}
\end{align}
Throughout, \(s\) denotes proper time along the detector worldline, and
we write
\begin{equation}
 \mathcal W^{+}(s,s')
 =\Wight\!\left(x_{\mathrm D}(s),x_{\mathrm D}(s')\right)
 \label{eq:W-restriction-convention}
\end{equation}
for the restriction of the spacetime two-point function to that
worldline, as announced in Sec.~\ref{sec:conventions}.

For a Hadamard state, the restriction of the Wightman distribution to a
smooth timelike curve is well defined; this is a standard microlocal
result, and we refer to
Refs.~\cite{Radzikowski1996Microlocal,BrunettiFredenhagenKohler1996,
Fewster2000,Hormander2003Analysis} rather than reproducing the argument
here.  What we do use is its consequence: the restricted two-point
function has a short-distance singularity of a universal form,
\begin{equation}
 \mathcal W^{+}(\Delta s)
 =-\frac1{4\pi^{2}(\Delta s-\ii\epsilon)^{2}}
 +O(1),
 \label{eq:universal-singularity}
\end{equation}
the same for every timelike curve and every Hadamard state. 

Smooth switching describes a finite interaction without the artificial
ultraviolet transients of abrupt switching and makes the response
well-defined before an infinite-duration limit is considered
\cite{Schlicht2004Unruh,LoukoSatz2006Clicks,
LoukoSatz2008Transition,Satz2007,SvaiterSvaiter1992Inertial,
SvaiterSvaiter1993Erratum}.  Once the
field state, physical trajectory,
switching, and gap are fixed, the result is independent of the
coordinate system and of the mode basis used to represent that same
state.  A different state can, of course, give a different
response.

If the state is stationary on the orbit, the restriction
\eqref{eq:W-restriction-convention} depends only on the proper-time
difference, and we may write
\begin{equation}
 \Wight(x_{\mathrm D}(s),x_{\mathrm D}(s'))
 =\mathcal W^+(\Delta s),
 \qquad \Delta s=s-s'.
 \label{eq:stationary-restriction}
\end{equation}

Choose the scaled switching
\(\chi_{\mathcal T}(s)=\chi(s/\mathcal T)\), divide the response by
\[
 \mathcal T\int_{-\infty}^{\infty}\chi(u)^2\dd u,
\]
and take \(\mathcal T\to\infty\).  The resulting
long-interaction-time transition rate is
\begin{equation}
 \dot{\mathcal F}(E)
 =\int_{-\infty}^{\infty}\dd\Delta s\,
 \e^{-\ii E\Delta s}\mathcal W^+(\Delta s).
 \label{eq:stationary-rate}
\end{equation}
It is the excitation rate when \(E>0\) and the de-excitation rate when
\(E<0\).
For stationary KMS correlators restricted to a worldline, scaled
adiabatic switching converges to
the stationary response and detailed-balance relation; a fixed finite
duration cannot reproduce that relation uniformly at arbitrarily large
gaps \cite{FewsterJuarezAubryLouko2016}.  Controlled long-time/small-gap
limits for circular motion further show that the order and scaling of
these limits matter \cite{ParryEtAl2025}.
It is worth noting that the detector gap is Fourier conjugate to
proper time, not to an arbitrary rotating coordinate time.

\subsection{Circular excitation in the Minkowski vacuum}
\label{sec:circular-excitation}

Consider the uniform circular trajectory
\begin{equation}
 x_{\mathrm D}(s)
 =(\gamma s,R,\varphi_0+\gamma\Omega_{\RR}s,z_0)
 \label{eq:circular-worldline}
\end{equation}
in inertial cylindrical coordinates.  It is at rest in the rigid chart
because it has fixed rigid coordinates
\((r,\varphi,z)=(R,\varphi_0,z_0)\), and it is stationary in the
geometric sense of Sec.~\ref{sec:conventions} because it follows an
orbit of the Killing field \(K_{\RR}\).  Here \(s\) is detector proper
time, \(R\) is the orbit radius, \(\Omega_{\RR}\) is the inertial angular
velocity, \(v\) is the tangential speed, \(\gamma\) is its Lorentz
factor, and \(\tau=T\) is rigid coordinate time, so
\[
 v=R\Omega_{\RR},\qquad
 \gamma=(1-v^2)^{-1/2},\qquad
 \Delta\tau=\gamma\Delta s.
\]
Its proper acceleration,
\begin{equation}
 a_{\mathrm c}=\frac{\gamma^2v^2}{R}
 =\gamma^2v\Omega_{\RR},
 \label{eq:circular-acceleration}
\end{equation}
sets the natural scale used in Fig.~\ref{fig:detector-response} below.

The mode-sum form of the rate follows in three steps.  First, restrict
the rigid Wightman function \eqref{eq:rigid-W-sum} to the worldline: on
Eq.~\eqref{eq:circular-worldline} the rigid spatial labels are fixed, so
\(\Delta\varphi=0\), \(\Delta z=0\), \(r=r'=R\), and the Bessel factor
becomes \(J_m^2(qR)\).  Second, use \(\Delta\tau=\gamma\Delta s\) to
express the remaining dependence through the proper-time difference, so
that the exponent of Eq.~\eqref{eq:rigid-W-sum} reduces to
\(-\ii\widetilde\omega\,\gamma\Delta s\), with \(\widetilde\omega\) the
corotating frequency \eqref{eq:corotating-frequency}.  Third, insert the
result into Eq.~\eqref{eq:stationary-rate} and perform the
\(\Delta s\) integral, which is the Fourier representation of a delta
function,
\begin{equation}
 \int_{-\infty}^{\infty}\dd\Delta s\,
 \e^{-\ii E\Delta s}\e^{-\ii\widetilde\omega\gamma\Delta s}
 =2\pi\,\delta\!\left(E+\gamma\widetilde\omega\right).
 \label{eq:delta-from-fourier}
\end{equation}
Collecting the factors gives the exact rate
\begin{equation}
 \begin{aligned}
 \dot{\mathcal F}_{\mathrm{circ}}(E)
 &=\sum_{m=-\infty}^{\infty}
 \int_0^\infty\dd q
 \int_{-\infty}^{\infty}\dd k\,
 \frac{qJ_m^2(qR)}{4\pi\omega}\\
 &\quad\times
 \delta\!\left[E+\gamma(\omega-m\Omega_{\RR})\right].
 \end{aligned}
 \label{eq:circular-mode-rate}
\end{equation}
The delta function in Eq.~\eqref{eq:circular-mode-rate} is the
statement of energy conservation for the detector--field exchange, with
the gap \(E\) balanced against \(\gamma\widetilde\omega\).  Its support
is the surface
\begin{equation}
 m\Omega_{\RR}=\omega+\frac E\gamma,
 \label{eq:excitation-condition}
\end{equation}
so a mode contributes only if Eq.~\eqref{eq:excitation-condition} has a
solution with \(\omega=\sqrt{q^2+k^2+\mu^2}\).  For \(E>0\) this forces
\(m\Omega_{\RR}>\omega\), that is
\(\widetilde\omega=\omega-m\Omega_{\RR}=-E/\gamma<0\): every
contributing mode has negative corotating frequency.
The detector can therefore excite even though
\(\beta^{\RR}=0\).  Negative values of the corotating frequency can satisfy the
detector's energy-conservation condition, while the same mode retains
positive inertial frequency and positive norm under the Klein--Gordon
product, as
established after Eq.~\eqref{eq:rigid-beta}.  The
detector transition rate probes whether the accelerated two-level
system exchanges energy with the field along its trajectory; the
Bogoliubov beta coefficient asks whether two mode decompositions use
different positive-frequency subspaces.  These are distinct questions.

The negative value of \(\omega-m\Omega_{\RR}\) is thus a negative
corotating energy.  It permits detector excitation by
the same kinematic energy bookkeeping that underlies rotational
superradiance, without implying that the present flat-spacetime
calculation contains an ergoregion or an amplified scattering wave
\cite{BekensteinSchiffer1998,MataczDaviesOttewill1993}.

The same calculation may be run explicitly in TT covering
coordinates.  A detector at rest in these coordinates at radius \(R\) has
\begin{equation}
 x_{\mathrm D}^{\TT}(s)
 =(t,r,\theta,z)=(s,R,\theta_0,z_0),
 \label{eq:TT-detector-orbit}
\end{equation}
because Eq.~\eqref{eq:TT-orbit} gives \(\dd s=\dd t\) on the orbit.

Write \(C_R=C(R)\) and \(S_R=S(R)\).  The inertial image coordinates
of two points on the orbit follow from Eq.~\eqref{eq:TT-inverse}:
\begin{align}
 T_{\mathrm D}(s)&=C_Rs+RS_R\theta_0,&
 \Phi_{\mathrm D}(s)&=C_R\theta_0+\frac{S_R}{R}s,
 \nonumber\\
 \Delta T_{\mathrm D}&=C_R\Delta s,&
 \Delta\Phi_{\mathrm D}&=\frac{S_R}{R}\Delta s.
 \label{eq:TT-detector-differences}
\end{align}
Consequently, restricting the TT mode sum
\eqref{eq:TT-W-sum} to Eq.~\eqref{eq:TT-detector-orbit} gives the
worldline correlation function
\begin{align}
 \mathcal W_{\TT}^{+}(\Delta s)
 &=\sum_{m=-\infty}^{\infty}\int_0^\infty\dd q
 \int_{-\infty}^{\infty}\dd k\,
 \frac{qJ_m^2(qR)}{8\pi^2\omega}
 \nonumber\\
 &\quad\times
 \exp\!\left[-\ii\left(
 \omega C_R-\frac{mS_R}{R}\right)\Delta s\right]
 \e^{-\omega\epsilon}.
 \label{eq:TT-worldline-Wightman}
\end{align}
The factor \(\e^{-\omega\epsilon}\) records the positive-inertial-frequency boundary value and is removed after taking the distributional
limit.  Substitution into the proper-time transform
\eqref{eq:stationary-rate} leaves one elementary Fourier integral:
\begin{align}
 &\int_{-\infty}^{\infty}\dd\Delta s\,
 \exp\!\left\{-\ii\left[
 E+\omega C_R-\frac{mS_R}{R}
 \right]\Delta s\right\}
 \nonumber\\
 &\qquad=2\pi\,
 \delta\!\left(E+\omega C_R-\frac{mS_R}{R}\right).
 \label{eq:TT-delta-from-fourier}
\end{align}
Multiplying the result in Eq.~\eqref{eq:TT-delta-from-fourier} by the mode-sum coefficient
\(q/(8\pi^2\omega)\) yields
\begin{align}
 \dot{\mathcal F}_{\TT}(E)
 &=\sum_m\int_0^\infty\dd q
 \int_{-\infty}^{\infty}\dd k\,
 \frac{qJ_m^2(qR)}{4\pi\omega}\nonumber\\
 &\quad\times
 \delta\!\left(E+\omega C_R-\frac{mS_R}{R}\right).
 \label{eq:TT-rate}
\end{align}
The support of this delta function is
\begin{equation}
 \frac{mS_R}{R}=\omega C_R+E,
 \label{eq:TT-excitation-condition}
\end{equation}
and for excitation, \(E>0\), it therefore requires
\(mS_R/R>\omega C_R\), the TT form of the negative-corotating-frequency
condition found after Eq.~\eqref{eq:excitation-condition}.
When the TT and rigid descriptions represent the same physical orbit,
\begin{equation}
 R\Omega_{\RR}=\tanh(\Omega_{\TT}R),\qquad
 \gamma=C_R,
 \label{eq:orbit-match}
\end{equation}
and hence
\begin{equation}
 \gamma\Omega_{\RR}=\frac{S_R}{R},\qquad
 \gamma(\omega-m\Omega_{\RR})
 =\omega C_R-\frac{mS_R}{R}.
 \label{eq:TT-rigid-frequency-match}
\end{equation}
The Fourier delta functions in Eqs.~\eqref{eq:circular-mode-rate} and
\eqref{eq:TT-rate} are therefore identical, including their proper-time
normalization, and the two rates coincide term by term.
This is the expected outcome, and it is worth stating plainly: the rigid
and TT coordinate descriptions assign different coordinate labels to the same circular
worldline in the same Minkowski vacuum, so they must predict the same
detector rate.  The agreement of
Eqs.~\eqref{eq:circular-mode-rate} and \eqref{eq:TT-rate} is a check of
the covariance established in Sec.~\ref{sec:covariance}, not an
independent physical result.

\subsection{The double pole and its Regularization}
\label{sec:renormalization}

Evaluating the massless Minkowski Wightman function on the circular
worldline \eqref{eq:circular-worldline} gives
\begin{equation}
 \mathcal W_{\mathrm{circ}}^+(s)
 =-\frac1{4\pi^2}
 \frac1{
 \gamma^2(s-\ii\epsilon)^2
 -4R^2\sin^2(\gamma\Omega_{\RR}s/2)
 },
 \label{eq:circular-W}
\end{equation}
where \(\epsilon\to0^+\) implements the positive-inertial-frequency
boundary value.  At \(s=0\), Eq.~\eqref{eq:circular-W} has the
universal Hadamard double pole identified after
Eq.~\eqref{eq:response}
\cite{Radzikowski1996Microlocal,Schlicht2004Unruh,Satz2007}.
Equation~\eqref{eq:circular-W} is a distributional boundary value and
not a function: the limit \(\epsilon\to0^{+}\) must be taken after the
Fourier transform.  Evaluating it at a fixed small \(\epsilon\) and
integrating numerically yields a regulator-dependent answer that does
not converge to the correct rate
\cite{Schlicht2004Unruh,Satz2007}.  We therefore cancel the double pole
analytically, before any numerical work.

For comparison, a detector at rest in an inertial frame may be
placed on \(x_{\mathrm{in}}(s)=(s,0,0,0)\).  Along this worldline,
\begin{align}
 \mathcal W_{\mathrm{in}}^+(s)
 &=-\frac1{4\pi^2(s-\ii\epsilon)^2}.
 \label{eq:inertial-worldline-W}
\end{align}
With the Fourier convention in Eq.~\eqref{eq:stationary-rate}, its
transform is
\begin{align}
 \dot{\mathcal F}_{\mathrm{in}}(E)
 &=\lim_{\epsilon\to0^+}
 -\frac1{4\pi^2}\int_{-\infty}^{\infty}
 \frac{\e^{-\ii Es}}{(s-\ii\epsilon)^2}\,\dd s
 \nonumber\\
 &=-\frac{E}{2\pi}\Theta(-E).
 \label{eq:inertial-rate}
\end{align}
For \(E>0\) the contour closes in the lower half-plane and encloses no
pole.  For \(E<0\) it closes in the upper half-plane and encloses the
double pole at \(s=\ii\epsilon\), giving
\(-E/(2\pi)\).

For numerical work we subtract this inertial Wightman function at the
same proper-time separation:
\begin{align}
 \Delta\mathcal W(s)
 &=\mathcal W_{\mathrm{circ}}^+(s)
 -\mathcal W_{\mathrm{in}}^+(s).
 \label{eq:Hadamard-subtraction}
\end{align}
The subtrahend is legitimate because
Eq.~\eqref{eq:inertial-worldline-W} carries exactly the universal
singularity \eqref{eq:universal-singularity}.  That singularity is the
same for every timelike curve and every Hadamard state, which is why one
and the same subtraction serves both the linearly accelerated and the
circular case.
For \(s\neq0\), this is the ordinary smooth function
\begin{equation}
 \Delta\mathcal W(s)
 =\frac1{4\pi^2}
 \left[
 \frac1{s^2}
 -\frac1{
 \gamma^2s^2-4R^2\sin^2(\gamma\Omega_{\RR}s/2)
 }\right].
 \label{eq:delta-W}
\end{equation}
Using the proper acceleration in
Eq.~\eqref{eq:circular-acceleration}, the behavior of this
Hadamard-subtracted correlation function near \(s=0\) is
\begin{equation}
 \Delta\mathcal W(s)
 =\frac{a_{\mathrm c}^2}{48\pi^2}+\order{s^2}.
 \label{eq:delta-W-limit}
\end{equation}
The double pole has cancelled analytically, leaving a finite
coincidence limit.

The full rate is recovered by adding the inertial transform,
\begin{equation}
 \dot{\mathcal F}_{\mathrm{circ}}(E)
 =-\frac{E}{2\pi}\Theta(-E)
 +2\int_0^\infty\dd s\,
 \cos(Es)\Delta\mathcal W(s).
 \label{eq:renormalized-rate}
\end{equation}
For excitation, \(E>0\), only the convergent integral remains.  For
decay (\(E<0\)), the first term restores inertial spontaneous emission.
This is Hadamard point splitting (or, in this flat setting, inertial
vacuum subtraction) and gives a finite numerical representation of
Eq.~\eqref{eq:stationary-rate}
\cite{LoukoSatz2008Transition,BarbadoVisser2012}.  It is not a change of
field state neither an ordinary normal ordering of the detector
response: the subtracted inertial distribution is added back through
Eq.~\eqref{eq:inertial-rate}.  The operation is a numerically convenient
representation of the original distributional Fourier transform.

\begin{figure*}[t]
 \centering
 \includegraphics[width=0.98\textwidth]{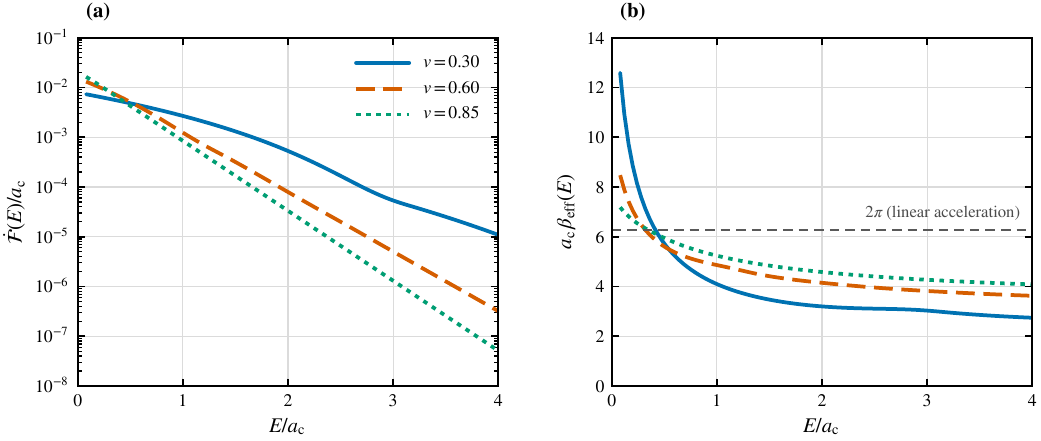}
 \caption{\label{fig:detector-response}
 Massless circular-detector response in the Minkowski vacuum, at fixed
 proper acceleration \(a_{\mathrm c}\), computed from the
 Hadamard-subtracted expression \eqref{eq:renormalized-rate}.
 (a)~The excitation rate \(\dot{\mathcal F}(E)/a_{\mathrm c}\), as a
 function of \(E/a_{\mathrm c}\), is nonzero for \(E>0\), but it is not
 a Planck spectrum.  (b)~The gap-dependent inverse temperature
 \(a_{\mathrm c}\beta_{\mathrm{eff}}(E)\), with
 \(\beta_{\mathrm{eff}}(E)=E^{-1}
 \ln[\dot{\mathcal F}(-E)/\dot{\mathcal F}(E)]\).
 A KMS state would give a horizontal curve; \(2\pi\), the value for
 linear acceleration at the same proper acceleration, is shown only as
 a reference.  Line color and dashing both identify the tangential
 speed.}
\end{figure*}

Figure~\ref{fig:detector-response} is obtained from
Eq.~\eqref{eq:renormalized-rate} with \(a_{\mathrm c}=1\), so that a chosen
tangential speed fixes
\[
 R=\frac{\gamma^2v^2}{a_{\mathrm c}},\qquad
 \Omega_{\RR}=\frac{a_{\mathrm c}}{\gamma^2v},
\]
using the accompanying reproducibility
package~\cite{RotatingVacuumCode}. The package includes the
figure-generation code and a notebook presenting the symbolic derivations
and numerical consistency checks in manuscript order.  The notebook provides a symbolic check of most calculations done in the text.  The correlation function calculation uses the Hadamard point splitting and an analytic leading-tail
correction; its parameters and convergence checks are given in
Appendix~\ref{app:numerics}.

\subsection{Why the response is not thermal}
\label{sec:not-thermal}

Nonzero stationary excitation is not enough to establish a
temperature.  A thermal state must satisfy the KMS condition stated in
Eq.~\eqref{eq:KMS-condition}, and detector rates must obey detailed
balance at one gap-independent inverse temperature
\cite{FullingRuijsenaars1987,Sewell1982ThermalStates,
Takagi1986VacuumNoise,FewsterJuarezAubryLouko2016}. 
Figure~\ref{fig:detector-response} shows directly that
\begin{equation}
 \beta_{\mathrm{eff}}(E)
 =\frac1E\ln\frac{\dot{\mathcal F}(-E)}
 {\dot{\mathcal F}(E)}
 \label{eq:beta-effective}
\end{equation}
depends on \(E\).

Here \(\beta_{\mathrm{eff}}(E)\) is only a spectral
detailed-balance diagnostic.  A detector can approach an asymptotic
state determined by its excitation and de-excitation rates without the
field state being KMS, and a gap-dependent ratio is not a thermodynamic
state variable \cite{JuarezAubryMoustos2019,GoodEtAl2020}.  This
distinction also separates response suppression from a genuine decrease
of detailed-balance temperature in discussions of anti-Unruh behavior
\cite{GarayMartinMartinezRamon2016}.

We also present a short analytic proof for the massless correlator.
Write Eq.~\eqref{eq:circular-W} in the form
\(\mathcal W^{+}_{\mathrm{circ}}=1/(4\pi^{2}D)\), and use the identity
\(2\sin^{2}(x/2)=(1-\cos x)\).  Then the resulting \(D\) is
\begin{equation}
 D(z)=-\gamma^2z^2+
 2R^2[1-\cos(\gamma\Omega_{\RR}z)].
 \label{eq:circular-denominator}
\end{equation}
It is the invariant separation \(\rho_\epsilon\) of
Eq.~\eqref{eq:rho-inertial} restricted to the circular worldline
\eqref{eq:circular-worldline}, with \(\Delta T=\gamma z\) and
\(|\Delta\bm x|^{2}=2R^{2}[1-\cos(\gamma\Omega_{\RR}z)]\). For a stationary bosonic two-point function, the KMS condition shifts
the complex proper-time difference by \(-\ii\beta\) and reverses the
operator order.  In terms of this denominator, the corresponding
boundary values would therefore require
\(D(z-\ii\beta)=D(-z)\).  For real \(z\),
\begin{equation}
 \operatorname{Im}D(z-\ii\beta)
 =2\gamma^2\beta z
 -2R^2\sin(\gamma\Omega_{\RR}z)
 \sinh(\gamma\Omega_{\RR}\beta).
 \label{eq:KMS-obstruction}
\end{equation}
The first term grows linearly with $z$, whereas the second is bounded and
sinusoidal.  They cannot cancel for every real \(z\) at any
\(\beta>0\).  The massless circular Minkowski-vacuum correlator is
therefore not finite-temperature KMS.

The complex zeros of \(D(z)\) control the decay of
\(\dot{\mathcal F}(E)\) at large positive gap, through the singularities
of \(\mathcal W^{+}_{\mathrm{circ}}\) nearest the real axis: the
imaginary part of the nearest zero sets the exponential rate at which
the excitation rate falls off as \(E\) grows.  Residue
and asymptotic analyses of the circular spectrum were given in
Refs.~\cite{Mueller1995RotatingSpectrum,BellLeinaas1983Thermometers};
recent fixed-proper-acceleration results give the small-gap, large-gap,
and ultrarelativistic limits
\cite{BiermannEtAl2020CircularTemperature,ParryLouko2025}.  The results
of Refs.~\cite{BiermannEtAl2020CircularTemperature,ParryLouko2025} also
show why, in the limits just quoted, the often cited circular
``temperature'' is an asymptotic comparison rather than an exact Planck
temperature.

If a mirror is placed at \(R_0<|\Omega_{\RR}|^{-1}\), the condition
\eqref{eq:excitation-condition} can no longer be satisfied for
\(E>0\), as shown in Sec.~\ref{sec:rigid-obstruction}, and the delta
function argument in Eq.~\eqref{eq:circular-mode-rate} has no roots.  The
detector
then remains unexcited in the bounded rotating ground-state problem
\cite{DaviesDrayManogue1996RotatingVacuum,
Nicolaevici2001NullResponse}.  This difference comes from the physical
boundary together with its boundary condition, which restrict the
allowed mode spectrum so that the rotating frequencies are positive.
Notice that the  mere use of rotating coordinates does not cause the difference;
nor is discreteness of the radial mode spectrum by itself the essential
point.

The numerical and KMS statements above were made for a massless scalar
field in \(3+1\) dimensions.  Detector regularization and spectra depend
on spacetime dimension \cite{HodgkinsonLouko2012}, and circular
detailed-balance temperatures differ between \(3+1\) and
\(2+1\) dimensions \cite{BiermannEtAl2020CircularTemperature,
ParryLouko2025,ParryFewsterLouko2026}; we record the difference only in
passing, and refer to those works for the comparison itself.  A nonzero
field mass suppresses
finite-time circular
responses and can introduce additional scale dependence
\cite{PanEtAl2024}; it does not alter the coordinate-covariance argument
or restore a positive global corotating generator.

\section{Discussion and conclusions}
\label{sec:conclusion}

The central result is a separation of state, coordinates, and
trajectory.  Starting with the Minkowski vacuum, inertial cylindrical,
rigid, and TT modes related by the stated coordinate transformations
are the same positive-inertial-frequency solutions written in
different variables.  For these explicitly coordinate-rewritten
bases, \(\beta^{\RR}=\beta^{\TT}=0\) follows directly and means only
that the coordinate changes have not introduced a different
positive-frequency splitting.  The nontrivial consistency checks are
that the displayed modes are admissible and normalized on the same
geometric Cauchy surface, and that their complete two-point functions
obey
\begin{align}
 \Wight_{\RR}(y,y')
 &=\Wight_{\M}\!\left(F_{\RR}(y),F_{\RR}(y')\right),
 \nonumber\\
 \Wight_{\TT}(x,x')
 &=\Wight_{\M}\!\left(F_{\TT}(x),F_{\TT}(x')\right).
 \label{eq:conclusion-transformations}
\end{align}

The direct rigid--TT transformation, Eq.~\eqref{eq:direct-W-equivalence},
provides a direct covariance cross-check of these relations.  A convergent conclusion, obtained
perturbatively and independently, is reported in
Ref.~\cite{KuboniwaMameda2026Vacuum}.

Neither rotating coordinate-time construction selects a global
stationary scalar ground state on unbounded Minkowski space, and the
two failures are of different kinds.  For rigid rotation the generator
is Killing, so a conserved corotating Hamiltonian does exist; what fails
is its spectrum, since \(H-\Omega_{\RR}J_z\) is unbounded below on
unbounded space, and the causal character of its orbits, which cease to
be timelike at the speed-of-light cylinder.  The obstruction is thus
spectral and causal, and it is removed by changing the region, as the
mirror construction of Sec.~\ref{sec:rigid-obstruction} shows.  For TT
rotation the situation is reversed: the congruence remains timelike
everywhere, so no causal obstruction arises, but the generating vector
field is not Killing, by Eqs.~\eqref{eq:TT-Lie} and
\eqref{eq:Lie-function}, so there is no conserved
generator to have a spectrum in the first place, and the
constant-time surfaces are not global Cauchy surfaces.  This obstruction
 is not removed by restriction to any
 nonempty open radial region; on a single fixed-radius orbit, however,
 the TT flow is proportional to an orbit-dependent Killing field.
Neither failure implies that Minkowski space lacks other Hadamard states
or that canonical quantization without stationarity is impossible.

The no-ground-state statement here is for the bosonic scalar field and
should not be transferred unchanged to fermions, whose corotating
quantizations have a different spectral structure
\cite{Iyer1982DiracRotating,AmbrusWinstanley2014RotatingStates,
AmbrusWinstanley2016,AmbrusWinstanley2021Rotation}.

A cylindrical boundary inside the speed-of-light surface supplies
extra data and can make a rigid rotating ground or thermal state well
defined, provided a boundary condition is imposed and the resulting
rotating spectrum, that is, the spectrum of \(H-\Omega_{\RR}J_z\), is
positive.  This is a new boundary-value problem,
not a consequence of changing coordinates.

The rigid and TT descriptions can label the same physical
circular worldline after their angular velocities are matched as in
Eq.~\eqref{eq:orbit-match}.  That circle is an orbit of the corotating
Killing field \(\Xi_R\) of Eq.~\eqref{eq:one-orbit-Killing}, so the
Minkowski-vacuum detector response is stationary
and nonzero.  Hadamard subtraction cancels the universal Wightman
double pole and restores the known inertial contribution.  The
resulting spectrum does not obey either the analytic KMS test or
gap-independent detailed balance.  Rotation therefore gives vacuum
excitation without a thermal bath, and without a second vacuum
associated with rotating time evolution, in the sense established in
Sec.~\ref{sec:covariance}.  The procedure to follow is to specify the
physical state and
worldline first, calculate the Wightman function on that worldline, and
then choose whichever coordinates make the calculation convenient.

Finally, the pointlike scalar monopole is a model of detection rather
than a complete description of an atom or storage-ring electron.
Stationary atomic response and Lamb shifts
\cite{AudretschMullerHolzmann1995}, electromagnetic dipole coupling in
circular motion \cite{JinHuYu2014}, and spin polarization in storage
rings \cite{BellLeinaas1987} display analogous vacuum excitation but
different spectra and observables.  Derivative-coupling detectors
provide another distinct model \cite{JuarezAubryLouko2014}.  These
comparisons support the qualitative statement that circular
acceleration can produce vacuum excitation without establishing that
all rotating detectors share the scalar UDW spectrum.

\begin{acknowledgments}
S.N.J. acknowledges financial support from the Coordena\c{c}\~ao de
Aperfei\c{c}oamento de Pessoal de N\'ivel Superior--Brasil (CAPES),
Finance Code 001, Grant No.~88887.142182/2025-00.  C.A.D.Z. is partially supported by Conselho Nacional de Desenvolvimento Cient\'ifico e Tecnol\'ogico (CNPq) under the grant no.~305610/2025-2. C.A.D.Z. is also funded by Funda\c{c}\~{a}o Carlos Chagas Filho de Amparo \`a Pesquisa do Estado do Rio de Janeiro (Faperj) under Grant no.~E-26/201{.}447/2021 (Programa Jovem Cientista do Nosso Estado).

\textit{Declaration of generative-AI use.---}
In preparing this manuscript the authors used OpenAI Codex (GPT-5.6 Sol)
and Anthropic Claude Code (Claude Opus 5) for language and structural
revision, for literature organization, and for assistance with the
\LaTeX{} and with the numerical implementation, including the code that
produces the figures. The authors take full responsibility for the scientific
content and the final form of the manuscript.
\end{acknowledgments}

\section*{Data availability}

The data that support the findings of this study, and the code that
generates them, are openly available in the reproducibility package of
Ref.~\cite{RotatingVacuumCode}.  That package contains the implementation
of Eq.~\eqref{eq:renormalized-rate}, the response values plotted in
Fig.~\ref{fig:detector-response}, the numerical error budget reported in
Appendix~\ref{app:numerics}, and a notebook of symbolic derivations and
numerical consistency checks of the equations presented here.  No other
data were generated or analyzed in this study.

\appendix

\section{Evaluation of the inertial Wightman mode sum}
\label{app:graf}

This appendix evaluates the Minkowski-vacuum Wightman function of
Sec.~\ref{sec:inertial-primer} in two independent ways and shows that
they agree.  We first perform the Cartesian momentum integral
\eqref{eq:cartesian-W-integral} directly, obtaining the closed forms
\eqref{eq:massless-W} and \eqref{eq:massive-W}; we then return to the
cylindrical mode sum \eqref{eq:inertial-mode-sum} and show that Graf's
theorem \eqref{eq:graf} reduces it to the same integral.

So that the calculation is self-contained, we repeat the starting point,
Eq.~\eqref{eq:cartesian-W-integral},
\begin{equation}
 \Wight_{\M,\mu}(X,X')
 =\int\frac{\dd^3\bm p}{(2\pi)^3\,2\omega_{\bm p}}\,
 \e^{-\ii\omega_{\bm p}(\Delta T-\ii\epsilon)
 +\ii\bm p\cdot\Delta\bm x},
 \label{eq:app-graf-start}
\end{equation}
with \(\omega_{\bm p}=\sqrt{\bm p^{2}+\mu^{2}}\).  For the massless
field, \(\mu=0\), so that \(\omega_{\bm p}=p\equiv|\bm p|\); set
\(\tau=\Delta T-\ii\epsilon\) and
\(d=|\Delta\bm X|\).  Let \(\vartheta\) denote the polar angle between
the momentum \(\bm p\) and the separation \(\Delta\bm X\), and let
\(\dd\Omega=\sin\vartheta\,\dd\vartheta\,\dd\varphi_{\bm p}\) be the
corresponding solid-angle element in momentum space, so that
\(\dd^{3}\bm p=p^{2}\dd p\,\dd\Omega\).  Performing the $\varphi_{\bm p}$ integral
in Eq.~\eqref{eq:app-graf-start} gives
\begin{align}
 \Wight_{\M,0}
 &=\frac1{(2\pi)^3}
 \int_0^\infty\frac{p^2\dd p}{2p}\,
 \e^{-\ii p\tau}
 \int\dd\Omega\,\e^{\ii pd\cos\vartheta}
 \nonumber\\
 &=\frac1{4\pi^2d}
 \int_0^\infty\dd p\,\sin(pd)\e^{-\ii p\tau}
 \nonumber\\
 &=\frac1{8\pi^2\ii d}
 \left[
 \frac1{\ii(\tau-d)}-\frac1{\ii(\tau+d)}
 \right]
 \nonumber\\
 &=\frac1{4\pi^2[d^2-\tau^2]}.
 \label{eq:massless-Fourier-app}
\end{align}
The negative imaginary part of \(\tau\) makes both exponential
integrals convergent and fixes their boundary values.  Since
\(d^2-\tau^2=\rho_\epsilon\), this is Eq.~\eqref{eq:massless-W}.

For \(\mu>0\) the same object, Eq.~\eqref{eq:app-graf-start}, is most
easily evaluated after Euclidean continuation.  Setting
\(T=-\ii T_E\) in the spacetime arguments of
Eq.~\eqref{eq:app-graf-start} and correspondingly \(p^{0}=\ii p_{4}\) in
its momentum variables, the Euclidean four-momentum is
\(p_E=(p_4,\bm p)\) with measure \(\dd^4p_E=\dd p_4\,\dd^3\bm p\), and
\(\Delta X_E\) is the Euclidean separation with
\(\rho_E=|\Delta X_E|\).  The continued two-point function is the
Euclidean propagator \(G_E(\rho_E)\) written below, and to evaluate it
we use the Schwinger representation 
\begin{equation}
 \frac1A=\int_0^\infty\dd\alpha\,\e^{-\alpha A},
 \qquad A>0,
 \label{eq:schwinger-rep-app}
\end{equation}
and the Gaussian momentum integral
\begin{equation}
 \int\frac{\dd^4p_E}{(2\pi)^4}
 \e^{-\alpha p_E^2+\ii p_E\cdot\Delta X_E}
 =\frac{1}{16\pi^2\alpha^2}
 \e^{-\rho_E^2/4\alpha},
 \label{eq:gaussian-app}
\end{equation}
one finds
\begin{align}
 G_E(\rho_E)
 &=\int\frac{\dd^4p_E}{(2\pi)^4}
 \frac{\e^{\ii p_E\cdot\Delta X_E}}{p_E^2+\mu^2}
 \nonumber\\
 &=\int_0^\infty\dd\alpha\,\e^{-\alpha\mu^2}
 \int\frac{\dd^4p_E}{(2\pi)^4}
 \e^{-\alpha p_E^2+\ii p_E\cdot\Delta X_E}
 \nonumber\\
 &=\frac1{16\pi^2}\int_0^\infty
 \frac{\dd\alpha}{\alpha^2}
 \exp\!\left[-\alpha\mu^2-\frac{\rho_E^2}{4\alpha}\right]
 \nonumber\\
 &=\frac{\mu}{4\pi^2\rho_E}K_1(\mu\rho_E).
 \label{eq:massive-Schwinger-app}
\end{align}
The final step uses the standard integral representation of the modified
Bessel function \cite{GradshteynRyzhik2014,DLMF}.
Analytic continuation with the same \(\ii\epsilon\) prescription as in
Eq.~\eqref{eq:cartesian-W-integral}
replaces \(\rho_E^2\) by \(\rho_\epsilon\) and yields
Eq.~\eqref{eq:massive-W}.

We now return to the cylindrical representation.  After using
Eq.~\eqref{eq:graf} to evaluate the azimuthal sum in
Eq.~\eqref{eq:inertial-mode-sum}, one finds by direct substitution
\begin{align}
 \Wight_{\M,\mu}
 =\int_0^\infty\dd q\int_{-\infty}^{\infty}\dd k\,
 \frac{q}{8\pi^2\sqrt{q^2+k^2+\mu^2}}\,
 J_0(qd_\perp)
 \nonumber\\
 {}\times
 \e^{\ii k\Delta z
 -\ii\sqrt{q^2+k^2+\mu^2}(\Delta T-\ii\epsilon)}.
 \label{eq:reduced-mode-sum-app}
\end{align}
To see directly how this matches the Cartesian integral \eqref{eq:app-graf-start}, decompose the momentum $\bm p = (p_x, p_y, p_z)$ into transverse and axial components by setting $p_z = k$ and parameterizing the transverse plane $(p_x, p_y)$ in polar coordinates $(q, \alpha)$, with $q = \sqrt{p_x^2 + p_y^2} \ge 0$. The momentum measure becomes $\dd^3\bm p = q\,\dd q\,\dd\alpha\,\dd k$, and the frequency is $\omega_{\bm p} = \sqrt{q^2 + k^2 + \mu^2}$. Choosing the reference axis for $\alpha$ along the transverse spatial separation $(\Delta x, \Delta y)$—whose magnitude is $\sqrt{(\Delta x)^2 + (\Delta y)^2} = d_\perp$—the exponent decomposes as
\[
 \bm p \cdot \Delta\bm x = q d_\perp \cos\alpha + k\Delta z.
\]
The angular integral over $\alpha$ in Eq.~\eqref{eq:app-graf-start} then evaluates to the Bessel representation
\[
 \int_0^{2\pi}\e^{\ii qd_\perp\cos\alpha}\dd\alpha = 2\pi J_0(qd_\perp),
\]
which immediately converts the Cartesian momentum integral \eqref{eq:app-graf-start} into the cylindrical mode sum result \eqref{eq:reduced-mode-sum-app}. This confirms that the cylindrical and Cartesian mode representations define the exact same two-point function.

Substituting the rigid or TT image coordinates, defined in Eqs.~\eqref{eq:rigid-transformation} and \eqref{eq:TT-inverse} respectively, into the invariant separation of the Wightman function yields their corresponding expressions \eqref{eq:rigid-W}--\eqref{eq:rigid-rho} and \eqref{eq:TT-W}--\eqref{eq:TT-rho} directly, without recomputing the $(q,k)$ momentum integrals.

\section{Wave Equation in TT Coordinates}
\label{app:TT-operator}

This appendix obtains Eq.~\eqref{eq:TT-KG} from the covariant
d'Alembertian \eqref{eq:box-covariant},
\[
 \Box\phi=\frac1{\sqrt{-g}}\,
 \partial_\mu\!\left(\sqrt{-g}\,g^{\mu\nu}\partial_\nu\phi\right),
\]
evaluated with the TT metric \eqref{eq:TT-metric}.  This is the same route
followed for the rigid chart in Eq.~\eqref{eq:rigid-box-covariant}, and
it establishes Eq.~\eqref{eq:TT-KG} as the wave operator of the geometry
written in Eqs.~\eqref{eq:TT-metric}--\eqref{eq:TT-metric-functions}.  Throughout we
order the coordinates as \(x^\mu=(t,r,\theta,z)\), and we recall that
\(C\), \(S\) and \(\eta=\Omega_{\TT}r\) of Eq.~\eqref{eq:TT-CS} are
functions of \(r\) alone, whereas \(\mathfrak A\), \(\mathfrak B\) and
\(\mathfrak Q\) of Eq.~\eqref{eq:TT-metric-functions} depend on \(t\)
and \(\theta\) as well.

Reading the components
off Eq.~\eqref{eq:TT-metric},
\begin{equation}
 g_{\mu\nu}=
 \begin{pmatrix}
 1 & \mathfrak A & 0 & 0\\
 \mathfrak A & -(1+\mathfrak Q) & \mathfrak B & 0\\
 0 & \mathfrak B & -r^{2} & 0\\
 0 & 0 & 0 & -1
 \end{pmatrix}.
 \label{eq:TT-metric-matrix-app}
\end{equation}
The whole calculation rests on one algebraic identity satisfied by the
functions \eqref{eq:TT-metric-functions}, namely
\begin{equation}
 \mathfrak Q=\frac{\mathfrak B^{2}}{r^{2}}-\mathfrak A^{2}.
 \label{eq:TT-Q-identity-app}
\end{equation}
The determinant of the metric in Eq.~\eqref{eq:TT-metric-matrix-app} can be computed from the determinant of the upper \((t,r,\theta)\) block, with a minus sign from $g_{zz}$:
\begin{align}
 \det g
 &=-\left[r^{2}(1+\mathfrak Q)-\mathfrak B^{2}
 +r^{2}\mathfrak A^{2}\right]\nonumber\\
 &=-r^{2},
 \label{eq:TT-determinant-app}
\end{align}
where Eq.~\eqref{eq:TT-Q-identity-app} provided the cancelations.  Hence
\begin{equation}
 \sqrt{-g}=r,
 \label{eq:TT-measure-app}
\end{equation}
the same measure factor as in the inertial cylindrical chart: the boost
functions cancel out of the determinant, and \(\mathfrak A\),
\(\mathfrak B\), \(\mathfrak Q\) do not enter the volume element at all.

Inverting
Eq.~\eqref{eq:TT-metric-matrix-app} gives
\begin{equation}
 g^{\mu\nu}=
 \begin{pmatrix}
 1-\mathfrak A^{2} & \mathfrak A
 & \dfrac{\mathfrak A\mathfrak B}{r^{2}} & 0\\[4pt]
 \mathfrak A & -1 & -\dfrac{\mathfrak B}{r^{2}} & 0\\[4pt]
 \dfrac{\mathfrak A\mathfrak B}{r^{2}}
 & -\dfrac{\mathfrak B}{r^{2}}
 & -\dfrac1{r^{2}}-\dfrac{\mathfrak B^{2}}{r^{4}} & 0\\[4pt]
 0 & 0 & 0 & -1
 \end{pmatrix}.
 \label{eq:TT-inverse-metric-app}
\end{equation}
Checking \(g^{\mu\lambda}g_{\lambda\nu}=\delta^{\mu}{}_{\nu}\) is
immediate and also uses Eq.~\eqref{eq:TT-Q-identity-app}.

The reason for writing
Eq.~\eqref{eq:TT-inverse-metric-app} in terms of \(\mathfrak A\) and
\(\mathfrak B\) is that the four components of
\(g^{\mu\nu}\partial_\nu\) collapse onto the single modified radial
derivative \(\mathfrak D_r\) of Eq.~\eqref{eq:TT-Dr}.  Acting on a
scalar \(\phi\),
\begin{align}
 g^{t\nu}\partial_\nu\phi
 &=\partial_t\phi+\mathfrak A\,\mathfrak D_r\phi,\nonumber\\
 g^{r\nu}\partial_\nu\phi
 &=-\mathfrak D_r\phi,\nonumber\\
 g^{\theta\nu}\partial_\nu\phi
 &=-\frac1{r^{2}}\partial_\theta\phi
 -\frac{\mathfrak B}{r^{2}}\,\mathfrak D_r\phi,\nonumber\\
 g^{z\nu}\partial_\nu\phi&=-\partial_z\phi.
 \label{eq:TT-contracted-gradient-app}
\end{align}

Inserting
Eqs.~\eqref{eq:TT-measure-app} and
\eqref{eq:TT-contracted-gradient-app} into
Eq.~\eqref{eq:box-covariant}.  Since \(r\) does not depend on \(t\),
\(\theta\) or \(z\), the factor \(\sqrt{-g}=r\) passes through those
three derivatives and cancels with the prefactor $1/\sqrt{-g}$, leaving
\begin{align}
 \Box_{\TT}\phi
 ={}&\partial_t\!\left(\partial_t\phi
 +\mathfrak A\,\mathfrak D_r\phi\right)
 -\frac1r\partial_r\!\left(r\,\mathfrak D_r\phi\right)
 \nonumber\\
 &+\partial_\theta\!\left(
 -\frac1{r^{2}}\partial_\theta\phi
 -\frac{\mathfrak B}{r^{2}}\,\mathfrak D_r\phi\right)
 -\partial_z^{2}\phi.
 \label{eq:TT-assembly-raw-app}
\end{align}
Carrying out the outer derivatives and separating the terms in which
they act on \(\mathfrak A\) and \(\mathfrak B\) from those in which they
act on \(\mathfrak D_r\phi\),
\begin{align}
 \Box_{\TT}\phi
 ={}&\partial_t^{2}\phi
 -\frac1r\partial_r\!\left(r\,\mathfrak D_r\phi\right)
 -\frac1{r^{2}}\partial_\theta^{2}\phi
 -\partial_z^{2}\phi
 \nonumber\\
 &+\left(\partial_t\mathfrak A
 -\frac1{r^{2}}\partial_\theta\mathfrak B\right)\mathfrak D_r\phi
 \nonumber\\
 &+\mathfrak A\,\partial_t\!\left(\mathfrak D_r\phi\right)
 -\frac{\mathfrak B}{r^{2}}\,
 \partial_\theta\!\left(\mathfrak D_r\phi\right).
 \label{eq:TT-assembly-app}
\end{align}

From
Eq.~\eqref{eq:TT-metric-functions},
\begin{equation}
 \partial_t\mathfrak A=\frac{S^{2}}{r},
 \qquad
 \frac1{r^{2}}\partial_\theta\mathfrak B
 =\frac{rS^{2}}{r^{2}}=\frac{S^{2}}{r},
 \label{eq:TT-divergence-identity-app}
\end{equation}
so that
\begin{equation}
 \partial_t\mathfrak A
 -\frac1{r^{2}}\partial_\theta\mathfrak B=0.
 \label{eq:TT-divergence-zero-app}
\end{equation}

The second line of Eq.~\eqref{eq:TT-assembly-app} vanishes by
Eq.~\eqref{eq:TT-divergence-zero-app}.  The third line supplies exactly
the missing pieces of a second \(\mathfrak D_r\), because
\begin{equation}
 \frac1r\mathfrak D_r\!\left(r\,\mathfrak D_r\phi\right)
 =\frac1r\partial_r\!\left(r\,\mathfrak D_r\phi\right)
 -\mathfrak A\,\partial_t\!\left(\mathfrak D_r\phi\right)
 +\frac{\mathfrak B}{r^{2}}\,
 \partial_\theta\!\left(\mathfrak D_r\phi\right),
 \label{eq:TT-Dr-outer-app}
\end{equation}
again because \(r\) is independent of \(t\) and \(\theta\).  Comparing
Eq.~\eqref{eq:TT-Dr-outer-app} with Eq.~\eqref{eq:TT-assembly-app}
gives
\begin{equation}
 \Box_{\TT}\phi
 =\partial_t^{2}\phi
 -\frac1r\mathfrak D_r\!\left(r\,\mathfrak D_r\phi\right)
 -\frac1{r^{2}}\partial_\theta^{2}\phi
 -\partial_z^{2}\phi,
 \label{eq:TT-box-final-app}
\end{equation}
and adding \(\mu^{2}\phi\) reproduces Eq.~\eqref{eq:TT-KG}.

\section{Klein--Gordon normalization of the TT modes}
\label{app:TT-KG-normalization}

This appendix supplies the normalization calculation summarized in
Eq.~\eqref{eq:TT-normalization}.  The essential point is to evaluate the
covariant Klein--Gordon product on a genuine Cauchy surface.  We choose
the inertial surface \(\Sigma_{T_0}:T=T_0\), whose equation in TT
coordinates is
\begin{equation}
 C(r)t+rS(r)\theta=T_0.
 \label{eq:TT-Cauchy-surface-app}
\end{equation}

Although the modes are displayed in TT covering coordinates in
Eq.~\eqref{eq:TT-modes}, they are the inertial cylindrical modes
expressed in those coordinates.  We may therefore parameterize the same geometric
surface by \((r,\Phi,z)\).  Its future-directed surface element is
\(\dd\Sigma^a=(\partial_T)^a r\,\dd r\,\dd\Phi\,\dd z\), and hence
we write \(\lambda=(q,m,k)\) and
\begin{align}
 \KG{u^{\TT}_{\lambda}}{u^{\TT}_{\lambda'}}
 ={}&\ii\int_0^\infty r\,\dd r
 \int_0^{2\pi}\dd\Phi
 \int_{-\infty}^{\infty}\dd z
 \nonumber\\
 &\times
 u^{\TT *}_{qmk}
 \overleftrightarrow{\partial_T}
 u^{\TT}_{q'm'k'} .
 \label{eq:TT-KG-product-app}
\end{align}
Here
\(f^*\overleftrightarrow{\partial_T}g
=f^*\partial_Tg-g\partial_Tf^*\).  On \(\Sigma_{T_0}\), the scalar
value of the TT mode is
\begin{align}
 u^{\TT}_{qmk}\big|_{\Sigma_{T_0}}
 &=\mathcal N_{q\omega}J_m(qr)
 \e^{\ii m\Phi+\ii kz-\ii\omega T_0},
 \nonumber\\
 \mathcal N_{q\omega}&=\frac1{2\pi}
 \sqrt{\frac{q}{2\omega}}.
 \label{eq:TT-mode-on-Cauchy-app}
\end{align}
Because \(\partial_Tu^{\TT}_{qmk}=-\ii\omega
u^{\TT}_{qmk}\), substitution into
Eq.~\eqref{eq:TT-KG-product-app} gives
\begin{align}
 \KG{u^{\TT}_{\lambda}}{u^{\TT}_{\lambda'}}
 ={}&(\omega+\omega')\mathcal N_{q\omega}
 \mathcal N_{q'\omega'}
 \e^{\ii(\omega-\omega')T_0}
 \nonumber\\
 &\times\int_0^\infty r\,\dd r\,
 J_m(qr)J_{m'}(q'r)
 \nonumber\\
 &\times\int_0^{2\pi}\dd\Phi\,
 \e^{\ii(m'-m)\Phi}
 \int_{-\infty}^{\infty}\dd z\,
 \e^{\ii(k'-k)z}.
 \label{eq:TT-KG-expanded-app}
\end{align}
The angular and longitudinal integrals, together with Bessel
orthogonality after \(m'=m\), are
\begin{align}
 \int_0^{2\pi}\dd\Phi\,
 \e^{\ii(m'-m)\Phi}&=2\pi\delta_{mm'},
 \nonumber\\
 \int_{-\infty}^{\infty}\dd z\,
 \e^{\ii(k'-k)z}&=2\pi\delta(k-k'),
 \nonumber\\
 \int_0^\infty r\,\dd r\,
 J_m(qr)J_m(q'r)&=\frac1q\delta(q-q').
 \label{eq:TT-orthogonality-integrals-app}
\end{align}
On the support of these delta functions, \(q'=q\), \(k'=k\), and
therefore \(\omega'=\omega\).  The remaining coefficient is
\begin{equation}
 (2\omega)\left(\frac{q}{8\pi^2\omega}\right)
 (2\pi)^2\frac1q=1,
 \label{eq:TT-normalization-coefficient-app}
\end{equation}
Define
\(\delta_{\lambda\lambda'}=
\delta(q-q')\delta_{mm'}\delta(k-k')\).  Equation
\eqref{eq:TT-normalization-coefficient-app} then proves
\begin{equation}
 \KG{u^{\TT}_{\lambda}}{u^{\TT}_{\lambda'}}
 =\delta_{\lambda\lambda'}.
 \label{eq:TT-KG-normalization-app}
\end{equation}
For the mixed product, the angular and longitudinal integrations instead
enforce \(m'=-m\) and \(k'=-k\).  The time-derivative coefficient is
proportional to \(\omega-\omega'\), which vanishes on the simultaneous
support of \(q'=q\) and \(k'=-k\).  Thus
\begin{align}
 \KG{u^{\TT}_{\lambda}}{u^{\TT *}_{\lambda'}}&=0,
 \nonumber\\
 \KG{u^{\TT *}_{\lambda}}{u^{\TT *}_{\lambda'}}
 &=-\delta_{\lambda\lambda'}.
 \label{eq:TT-mixed-normalization-app}
\end{align}
This appendix therefore verifies the TT modes' normalization and
their orthogonality to the conjugate modes on the inertial Cauchy
surface.  No normalization on a \(t=\mathrm{const}\) surface is used or
required.  Together with the defining identity
\eqref{eq:mode-pushback-identity}, these relations are consistent with
\(\alpha^{\TT}_{\lambda\lambda'}=\delta_{\lambda\lambda'}\) and
\(\beta^{\TT}_{\lambda\lambda'}=0\); the latter values do not constitute
an additional result of the normalization calculation.

\section{TT Wightman mode sum}
\label{app:TT-W-sum}

The exponent in the TT mode \eqref{eq:TT-modes} is
\begin{align}
 &\ii(mC-\omega rS)\theta
 -\ii\left(\omega C-\frac{mS}{r}\right)t+\ii kz
 \nonumber\\
 &\qquad
 =\ii m\left(C\theta+\frac Sr t\right)
 -\ii\omega(Ct+rS\theta)+\ii kz
 \nonumber\\
 &\qquad
 =\ii m\Phi_x-\ii\omega T_x+\ii kz,
 \label{eq:TT-phase-app}
\end{align}
where Eq.~\eqref{eq:TT-images} is used to obtain the last line.  The
conjugate
mode at \(x'\) supplies the opposite phase.  Consequently
\begin{align}
 u^{\TT}_{qmk}(x)u^{\TT *}_{qmk}(x')
 ={}&\frac{q}{8\pi^2\omega}J_m(qr)J_m(qr')
 \nonumber\\
 &\times\e^{\ii m(\Phi_x-\Phi_{x'})
 +\ii k(z-z')-\ii\omega(T_x-T_{x'})}.
 \label{eq:TT-mode-product-app}
\end{align}
Inserting the positive-frequency boundary value
\(T_x-T_{x'}\to T_x-T_{x'}-\ii\epsilon\) and summing over the complete
mode labels gives Eq.~\eqref{eq:TT-W-sum}.  Graf's identity then gives
\begin{equation}
 \sum_{m=-\infty}^{\infty}
 J_m(qr)J_m(qr')\e^{\ii m(\Phi_x-\Phi_{x'})}
 =J_0(qd_{\perp,\TT}),
 \label{eq:TT-Graf-app}
\end{equation}
where
\[
 d_{\perp,\TT}^2
 =r^2+r'^2-2rr'\cos(\Phi_x-\Phi_{x'}).
\]
The remaining \(q\) and \(k\) integrals are therefore the
Cartesian momentum integral evaluated at the inertial image points
\(F_{\TT}(x)\) and \(F_{\TT}(x')\), proving
Eqs.~\eqref{eq:TT-W}--\eqref{eq:TT-W-transformation}.

\section{Numerical Calculation of the Response Function}
\label{app:numerics}

This appendix details the regularized variables, series expansions, and
quadrature schemes used to evaluate the detector transition rate
\eqref{eq:renormalized-rate} in the numerical code that generates
Fig.~\ref{fig:detector-response}.

We express the integration in terms of the dimensionless proper time
$u=a_{\mathrm c}s$ and energy gap $e=E/a_{\mathrm c}$. With
$\gamma=(1-v^2)^{-1/2}$, the Hadamard-subtracted Wightman function
\eqref{eq:delta-W} becomes
\begin{equation}
 \frac{\Delta\mathcal W(u)}{a_{\mathrm c}^2}
 =\frac1{4\pi^2}
 \left[
 \frac1{u^2}
 -\frac1{
 \gamma^2u^2-4\gamma^4v^4
 \sin^2[u/(2\gamma v)]
 }\right].
 \label{eq:numerical-integrand}
\end{equation}
Direct floating-point evaluation of Eq.~\eqref{eq:numerical-integrand}
for small $u$ suffers from severe catastrophic cancellation: both
terms diverge as $u^{-2}$ as $u\to0$, while their difference approaches
the finite limit $1/(48\pi^2)$ given in Eq.~\eqref{eq:delta-W-limit}.
To eliminate this subtraction error, we define the variables
\begin{equation}
 x=\frac{u}{2\gamma v},\qquad
 N=4\gamma^4v^4\bigl(x^2-\sin^2x\bigr).
 \label{eq:numerical-stable-variables}
\end{equation}
This allows
Eq.~\eqref{eq:numerical-integrand} to be rewritten in the algebraically
equivalent form
\begin{equation}
 \frac{\Delta\mathcal W(u)}{a_{\mathrm c}^2}
 =\frac{N}{4\pi^2u^2(u^2+N)}.
 \label{eq:numerical-stable-integrand}
\end{equation}

For small arguments, evaluating $x^2-\sin^2x$ directly still incurs loss
of precision because $\sin x \approx x$. Using the half-angle identity
$x^2-\sin^2x = x^2-\frac12(1-\cos 2x)$ to linearize the trigonometric
power, we expand the numerator as
\begin{align}
 x^2-\sin^2x
 ={}&\frac{x^4}{3}-\frac{2x^6}{45}+\frac{x^8}{315}
 \nonumber\\
 &-\frac{2x^{10}}{14175}+\frac{2x^{12}}{467775}
 +\order{x^{14}}.
 \label{eq:numerical-series}
\end{align}
The code employs Eq.~\eqref{eq:numerical-series} for $|x| < 0.15$ and
switches to direct evaluation of $x^2-\sin^2x$ for $|x| \ge 0.15$.
The first omitted term in the series is $-4x^{14}/42567525$. At the
switching boundary $|x|=0.15$, the relative truncation error is
$1.6\times10^{-15}$, which is at the level of double-precision rounding
noise and safely below the precision loss ($\sim 10^{-14}$) incurred by
direct floating-point subtraction just above the threshold. At $u=0$,
the integrand is assigned its exact coincidence limit
$\Delta\mathcal W(0)/a_{\mathrm c}^2 = 1/(48\pi^2)$.

At large $u$, the bracket in Eq.~\eqref{eq:numerical-stable-integrand}
behaves as
\begin{equation}
 \frac{\Delta\mathcal W(u)}{a_{\mathrm c}^2}
 =\frac{v^2}{4\pi^2u^2}+\order{u^{-4}}.
 \label{eq:numerical-tail}
\end{equation}
The integral over $[0,\infty)$ is split at an upper cutoff $U$. The
interval $[0,U]$ is integrated numerically, while the leading
oscillatory $u^{-2}$ tail over $[U,\infty)$ is integrated analytically:
\begin{align}
 \int_U^\infty\frac{\cos(eu)}{u^2}\dd u
 &={}\frac{\cos(eU)}{U}
 -|e|\left[\frac{\pi}{2}-\operatorname{Si}(|e|U)\right],
 \label{eq:numerical-tail-integral}
\end{align}
where $\operatorname{Si}(y)=\int_0^y t^{-1}\sin t\,\dd t$ is the sine
integral.

The production curves in Fig.~\ref{fig:detector-response} use composite
24-point Gauss--Legendre quadrature \cite{DavisRabinowitz1984,DLMF}
partitioned into unit panels ($\Delta u = 1.0$) up to the cutoff
$U=800$, complemented by the analytic tail
\eqref{eq:numerical-tail-integral}. To verify convergence, the
quadrature was refined to 32-point Gauss--Legendre rules on panels of
width $\Delta u = 0.75$ with an extended cutoff $U=1200$. Across all 297
evaluation points, the maximum absolute and relative shifts between the
two schemes are
\[
 \max\frac{|\delta\dot{\mathcal F}|}{a_{\mathrm c}}
 =1.2\times10^{-12},\qquad
 \max\frac{|\delta\dot{\mathcal F}|}{|\dot{\mathcal F}|}
 =1.6\times10^{-7}.
\]
The largest relative difference occurs at the smallest plotted rate.

After subtracting the analytic tail \eqref{eq:numerical-tail}, the
neglected integrand is $\order{u^{-4}}$, giving an absolute cutoff
truncation error of $\order{U^{-3}}$. Bounding $\sin^2 x \le 1$ in
Eq.~\eqref{eq:numerical-stable-integrand} limits this residual
contribution to $6.9\times10^{-11}$ at $U=800$ across all plotted
velocities ($1.3\times10^{-3}$ relative to the smallest plotted rate).
For the zero-gap case $e=0$, successive doubling of the cutoff from
$U=100$ to $U=1600$ yields observed convergence orders of
$2.95 \le p \le 3.03$ for all three velocities, where
$p=\log_2(|R_U-R_{2U}|/|R_{2U}-R_{4U}|)$, in agreement with the
expected $\order{U^{-3}}$ asymptotic scaling.

As an independent validation, adaptive QUADPACK Fourier integration
(via SciPy) was applied directly to
Eq.~\eqref{eq:numerical-stable-integrand} on $[0,\infty)$ without panel
partitioning or tail separation. At five representative points spanning
$0.08 \le e \le 4.00$ across all three velocities, the maximum
discrepancy between QUADPACK and the composite Gauss--Legendre
quadrature is $1.5\times10^{-12}$ absolutely and $1.0\times10^{-7}$
relatively. This agreement between distinct numerical methods confirms
that numerical errors remain several orders of magnitude below the
graphical resolution of Fig.~\ref{fig:detector-response}.

Computations were carried out using Python 3.11.9, NumPy 2.4.6, SciPy
1.15.1, and Matplotlib 3.10.0; symbolic checks in the verification
notebook use SymPy 1.13.1 and mpmath 1.3.0. Complete environment
specifications and reproduction instructions accompany the archived
repository~\cite{RotatingVacuumCode}.

\setlength{\bibsep}{-1pt}
\bibliography{references}

\end{document}